\documentclass[twocolumn, secnumaRabic, amssymb, nobibnotes, aps, showpacs, superscriptaddress]{revtex4-1}
\usepackage{graphicx,amsmath,booktabs}%
\usepackage{color, array,amsmath,amsthm,amssymb,bm}
\usepackage[bookmarks=false]{hyperref}
\hypersetup{colorlinks=true,citecolor=blue,linkcolor=blue,urlcolor=blue,pdfstartview=FitH,bookmarksopen=true}
\begin{document}

\title{Rabi-error and Blockade-error-resilient All-Geometric Rydberg Quantum Gates}%

\author{S.-L. Su}
\author{Li-Na Sun}
\affiliation{School of Physics, Zhengzhou University, Zhengzhou 450001, China}
\author{B.-J. Liu}
\email[Email: ]{lbjie2010@163.com}
\altaffiliation[Present Address:~]{Department of Physics, University of Massachusetts-Amherst, Amherst, MA 01003, USA}
\affiliation{Department of Physics, Southern University of Science and Technology, Shenzhen, Guangdong 518055, China}

\author{L.-L. Yan}
\email[Email: ]{llyan@zzu.edu.cn}
\affiliation{School of Physics, Zhengzhou University, Zhengzhou 450001, China}
\author{M.-H. Yung}
\affiliation{Department of Physics, Southern University of Science and Technology, Shenzhen, Guangdong 518055, China}
 \author{W. Li}
\email[Email: ]{weibin.Li@nottingham.ac.uk}
\affiliation{School of Physics and Astronomy, University of Nottingham, Nottingham NG7 2RD, United Kingdom}
\author{M. Feng}
\affiliation{State Key Laboratory of Magnetic Resonance and Atomic and Molecular Physics, Wuhan Institute of Physics and Mathematics, Innovation Academy of Precision Measurement Science and Technology, Chinese Academy of Sciences, Wuhan 430071, China}
\affiliation{Research Center for Quantum Precision Measurement, Guangzhou Institute of Industry Technology, Guangzhou, 511458, China}
\affiliation{Department of Physics, Zhejiang Normal University, Jinhua 321004, China}
\date{\today}%
\begin{abstract}

We propose a nontrivial two-qubit gate scheme in which Rydberg atoms are subject to designed pulses resulting from geometric evolution processes. By utilizing a hybrid robust non-adiabatic and adiabatic geometric operations on the control atom and target atom, respectively, we improve the robustness of two-qubit Rydberg gate against Rabi control errors as well as blockade errors in comparison with the conventional two-qubit blockade gate. Numerical results with the current state-of-the-art experimental parameters corroborates the above mentioned robustness. We also evaluated the influence induced by the motion-induced dephasing  and the dipole-dipole interaction and imperfection excitation induced leakage errors, which both could decrease the gate fidelity. Our scheme provides a promising route towards systematic control error (Rabi error) as well as blockade error tolerant geometric quantum computation on neutral atom system. 

\end{abstract}
\maketitle

\section{introduction}

Neutral atoms have strong dipole-dipole interactions when excited to high-lying Rydberg states~\cite{TF:1994, MTK:2010,dp2010,lar2013}. The dipole-dipole-interaction induced Rydberg blockade has many important applications in quantum computation~\cite{DJP:2000,MMR:2001}. Experimentally, the Rydberg blockade has been observed~\cite{ett2009,ayt2009}, and furthermore, quantum CNOT gates as well as quantum entangled states~\cite{lex2010,xla2010,tac2010,kmt2015,ypx2017,Picken_2018,saa2018,levine2019parallel,graham2019rydberg} using Rydberg atoms have also been achieved. And the toric code topological order has also been predicted based on the Rydberg blockade~\cite{lukintoplogicalorder}. Quantum logic gates based on Rydberg blockade are often accompanied with blockade errors~\cite{saffmanblockadeerror} proportional to $(\Omega/V)^2$, where $\Omega$ and \emph{V} denote the Rabi frequency and Rydberg-Rydberg-interaction~(RRI)~strength, respectively. Although blockade errors can be reduced by increasing the RRI strength, the performance of the quantum computation scheme will be affected inevitably more or less since the mechanical effect would be increased due to the increase of RRI strength~\cite{wci2013}. The blockade error can be minimized through considering rational generalized Rabi frequency~\cite{shi2017free} or taking into consideration of dark-state dynamics that contain Rydberg states~\cite{klausdarkstate2017}. In addition to the blockade error, the control error, such as the Rabi frequency error induced by laser intensity fluctuations at high Rabi frequencies~\cite{kalerabifluctuation,*Madjarov2020}, is another resource of infidelity commonly encountered in the Rydberg quantum computation.

The Abelian geometric phase~\cite{Berry1984,Aharonov1987} and non-Abelian holonomy~\cite{ZANARDI1999,ANANDAN1988,Wilczek1984,sjovist2008} depend only on the global properties of the evolution trajectories of cyclic processes. On that basis, the geometric quantum logic gates based on Abelian and non-Abelian geometric phases (holonomy) are robust against local noises during the gate evolution~\cite{zhu2005,De2003,Leek2007,Filipp2009,Berger2013}. Earlier geometric quantum computation schemes usually rely on adiabatic processes that can suppress the transition between different instantaneous eigenstates of the Hamiltonian~\cite{Jones2000,duan2001science,wu2005,Wu2013adiabaticgeometric,duan2019adiabaticexperiment}. Nevertheless, since the adiabatic process requires longer evolution time to satisfy the adiabatic condition, the scheme may suffer from the influence of decoherence although it is robust to systematic control errors. Then, the nonadiabatic geometric quantum computation~\cite{Wang2001,Zhu2002,Thomas2011,Zhao2017,Chen2018} and nonadiabatic holonomic quantum computation (NHQC)~\cite{Sj_qvist_2012,xu2012} have been proposed to reduce the evolution time of geometric gates, which can enhance the robustness of the scheme on decoherence~\cite{Johansson2012,xue2015,Azimi2017,jingjun2017,Ramberg2019,zhao2020}. Experimentally, progresses in nuclear magnetic resonance system~\cite{Feng2013Experimental,Zhu2019Single}, superconducting qubits~\cite{Xu2020Experimental,Zhao2021Experimental,Song2017Continuous,han2020experimental,AbdumalikovJr2013Experimental,Danilin2018Experimental,Egger2019Entanglement} and nitrogen-vacancy centers in diamond~\cite{Zu2014Experimental,Nagata2018Universal,Arroyo-Camejo2014Room,Sekiguchi2017Optical,Zhou2017Holonomic,Ishida2018Universal} have confirmed the theoretical schemes. However, these nonadiabatic schemes are sensitive to the experimental control errors \cite{zheng2016}, which reduce the real usefulness of NGQC and NHQC. Recently, to overcome the problem, Liu et al \cite{liu2019} proposed a NHQC+ scheme by combining nonadiabatic geometric quantum computation with optimal control technology, but at the cost of complicated pulses and gate time \cite{Kang2020,Guo2020,fqguo2020,Yehong2021,Yan2019Experimental,Anmin2020,DongYang2021}. To balance all of speed, flexibility and robustness of geometric gates, the super-robust pulse geometric quantum computation scheme has been theoretically proposed \cite{liu2021super-robust} and experimentally realized \cite{li2021realization}.

In this paper, we employ the geometric processes to construct the nontrivial two-qubit Rydberg  gate under the consideration of the dark-state dynamics as described in Ref.~\cite{klausdarkstate2017}, where we can realize two-qubit gates robust to Rabi control as well as blockade errors by the hybrid of robust non-adiabatic and adiabatic geometric operations on the control atom and target atom. Through a thorough numerical analysis on the performance of our scheme and conventional Rydberg two-qubit scheme under current experimental conditions, the control error caused by the deviation of the laser Rabi frequency and the blockade error can be significantly suppressed. We also consider the motion-induced dephasing as well as dipole-dipole-interaction- and imperfection-excitation-induced leakage errors.  Our scheme is suitable and useful for Rydberg experimental platforms where some sever conditions, such as ultrastable Rabi frequency and very strong Rydberg atom interactions, can be relaxed.

\begin{figure}[tbp]
	\centering
	\includegraphics[width=8.5cm]{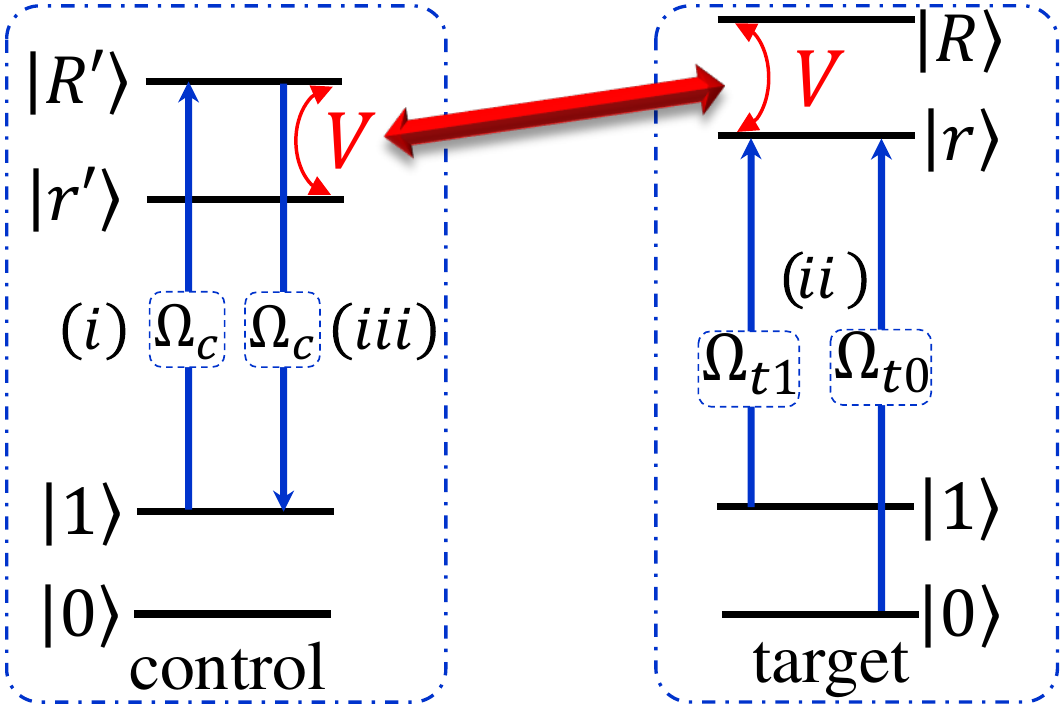}
	\caption{Energy levels of the control~(left) and target~(right) qubits to construct C{\footnotesize NOT} gate. $|0\rangle\equiv |6s_{1/2},~f=3,~m_f=0\rangle$ and $|1\rangle\equiv |6s_{1/2},~f=4,~m_f=0\rangle$ are two long-lived ground states of Cs atom clock states. $|R'\rangle\equiv|101s_{1/2},~m=1/2\rangle$ and $|r'\rangle\equiv|101p_{3/2},~m=3/2\rangle$ are Rydberg states of the control atom, and $|R\rangle\equiv|109p_{3/2},~m=3/2\rangle$ and $|r\rangle\equiv|109s_{1/2},~m=1/2\rangle$ are Rydberg states of the target atom. We consider the static Stark field $15.4~V/m$ directed along the quantization axis that makes the Rydberg pairs energy degenerate and leads to resonant interaction with $C_3=64.4$~GHz~$\cdot\mu m^3$~\cite{klausdarkstate2017}. $V$ denotes the RRI strength relevant to $C_3$ and the inter-atomic distance. For the control atom, the transition $|1\rangle\leftrightarrow|R'\rangle$ is driven resonantly by the laser with time-dependent Rabi frequency $\Omega_c(t)\equiv|\Omega_{c}(t)|e^{i\varphi_c(t)}$. For the target atom, the transition $|0\rangle(|1\rangle)\leftrightarrow|r\rangle$ is driven resonantly by the lasers with time-dependent Rabi frequency $\Omega_{t0}(t)[\Omega_{t1}(t)]$ and phase $\varphi_{t0}(\varphi_{t1})$.
	}\label{fig1}
\end{figure}

\section{MODEL}\label{sec2}
The protocol to achieve the two-qubit C{\footnotesize NOT} gate is based on the dark state scheme~\cite{klausdarkstate2017} and consists of the following three steps sketched in Fig.~\ref{fig1}. Step~(i) is to apply a resonate laser to achieve the geometric operation $|1\rangle\rightarrow|R'\rangle$ of the control atom. In the rotating wave approximation
and the interaction framework, the Hamiltonian of this step can be written as ($\hbar \equiv 1$)
\begin{eqnarray}\label{eq11}
H_c(t)&=&\frac{1}{2}\left(
\begin{array}{cc}
0 & \Omega_c^* \\
\Omega_c  & 0 \\
\end{array}
\right)
\end{eqnarray}
in the basis $\{|1\rangle,~|R'\rangle\}$ with the control parameters of the lasers $\Omega_c\equiv\Omega_{c}(t)=|\Omega_{c}(t)|e^{-i\varphi_{c}(t)}$.
In general, it is difficult to analytically solve the dynamical evolution $U(t)=\mathcal{T} e^{-i \int_{0}^{t} H_{c}(t^{\prime}) d t^{\prime}}$ with time-dependent Hamiltonian due to the time-ordering operator.

To achieve the robust geometric gates, we adopt the inverse engineering method~\cite{Dridi2020,DaemsPRL2013,liu2021coherent} by choosing a pair of states $|\phi_k(t)\rangle$ following the time-dependent Schr\"{o}dinger equation,
\begin{eqnarray}\label{eq5}
|\phi_1(t)\rangle&\equiv& e^{i\gamma}\left[\cos(\theta/2)|1\rangle -\sin(\theta/2)e^{-i\eta}|R'\rangle\right] ,
\cr\cr|\phi_2(t)\rangle&\equiv& e^{-i\gamma}\left[\sin(\theta/2)e^{i\eta}|1\rangle +\cos(\theta/2)|R'\rangle\right],
\end{eqnarray}
where $\gamma$, $\eta$, and $\theta$ are time-dependent parameters. Explicitly, we find that the control parameters of the laser are governed by the following coupled differential equations~[see Appendix \ref{appendixA}],
\begin{eqnarray}\label{e10}
\left|\Omega_{c}(t)\right|&=&\sqrt{\dot{\theta}^{2}+\dot{\eta}^{2} \tan ^{2} \theta},  \cr\cr
\varphi_{c}(t)&=&\eta-\frac{\pi}{2}-\arctan \left(\frac{\dot{\eta} \tan \theta}{\dot{\theta}}\right),  \cr\cr
\dot{\gamma}(t)&=&-\frac{\sin ^{2}(\theta / 2)}{\cos \theta} \dot{\eta}.
\end{eqnarray}
After a cyclic evolution, i.e., $|\phi_k(\tau)\rangle=\exp \left[i(-1)^{k} \gamma(\tau)\right]|\phi_k(0)\rangle~(k=1,~2)$, the acquired non-adiabatic geometric phase (Aharonov-Anandan phase)~\cite{Aharonov1987} is given by
\begin{equation}\label{EQY}
    \gamma_{g}=\Delta \gamma+\int_{0}^{\tau}\left\langle\phi_{1}(t)|H(t)| \phi_{1}(t)\right\rangle d t \ ,
\end{equation}
where $\Delta \gamma=\gamma(\tau)-\gamma(0)$ is the global phase, and the second part on the right-hand side of Eq. (\ref{EQY}) denotes the dynamical phase. To remove the dynamical phase, one simple choice is to satisfy the parallel transport condition, $\left\langle\phi_{k}(t)|H(t)| \phi_{k}(t)\right\rangle=0$. Specifically, we find that the control parameters need to satisfy the following condition
\begin{equation}\label{Condition1}
\dot{\eta}\sin\theta=0 \ .
\end{equation}
Then the resulting unitary evolution becomes purely geometric , i.e. 
$U(\tau)=e^{i \gamma_{g}}|\phi_{1}(0)\rangle\langle\phi_{1}(0)|+e^{-i \gamma_{g}}| \phi_{2}(0)\rangle\langle\phi_{2}(0)|$,
which is non-diagonal in the basis \{$|1\rangle, |R'\rangle$\},
\begin{equation}\label{Gate1}
    U(\tau)=e^{-i\gamma_{g} \boldsymbol{n} \cdot \sigma^{(c)}}
\end{equation}
where $\boldsymbol{n}=\left[\sin \theta(0) \cos \eta(0), -\sin \theta(0) \sin \eta(0), \cos \theta(0)\right]$, ${\sigma^{(c)}}=[\sigma_{x}^{(c)},~\sigma_{y}^{(c)},~\sigma_{z}^{(c)}]$ denotes the Pauli matrix of the control atom. Note that the robustness of geometric gate in Eq. (\ref{Gate1}) against the experimental Rabi error, i.e., $\Omega_{C} \rightarrow \Omega_{C}(1+\xi)$ with relative Rabi frequency deviation $\xi$, is no more advantage than standard dynamical gate  \cite{zheng2016}. To further enhance the robustness on the error, the additional dynamical effect between the states $|\phi_{1}(t)\rangle$ and $|\phi_{2}(t)\rangle$ should be eliminated \cite{liu2021super-robust}, i.e., $\int_{0}^{\tau}\left\langle\phi_{1}(t)|H(t)| \phi_{2}(t)\right\rangle d t=0$. Specifically, the control parameters should satisfy the following constrain,
\begin{equation}\label{e9}
\int_0^{\tau}\frac{\dot{\theta}}{2}\exp\left(-i\int_0^{t}\frac{\dot\eta}{\cos\theta}dt' \right)dt=0,
\end{equation}
where $\tau = \tau_1$ is the total time for step (i).

\begin{table}[h!]
\renewcommand\arraystretch{1.5}
\caption{One set of possible parameters of the laser amplitude and phase to implement the super-robust geometric quantum operations in step (i).}
\begin{tabular}{ cccc  }
 \hline
 \hline
  & $~t~\in$~[$0,~\frac{\tau_1}{3}$]~ & ~$t~\in$~($\frac{\tau_1}{3},~\frac{2\tau_1}{3}$]~ & $~t~\in$~($\frac{2\tau_1}{3},~\tau_1$]~\\
 \hline
 $\varphi_c(t)$	& $\frac{\pi}{3}$	& $-\frac{\pi}{3}$ & $\frac{\pi}{3}$\\
 \hline
 ~~$|\Omega_c(t)|$~~  & $\frac{3\pi}{\tau_1}$ & $\frac{3\pi}{\tau_1}$	& $\frac{3\pi}{\tau_1}$\\
   \hline\hline
 \end{tabular}\label{t1}
\label{symbols}
 \end{table}

During the step (i), we implement the $|1\rangle\rightarrow|R'\rangle$ operation on the control atom, which is equivalent to achieving the NOT gate ($U=\sigma_x$) when the initial state is $|1\rangle$. To satisfy the conditions in Eqs.~(\ref{Condition1}) and ~(\ref{e9}) for the robust NOT gate, one can set 
\begin{equation}\label{eq08}
|\Omega_c(t)|=\dot{\theta},~\varphi_{c} = \eta-\frac{\pi}{2}.
\end{equation}
A set of listed in Table.~\ref{t1}, which satisfy the constrain in Eqs.~(\ref{Condition1}) and~(\ref{e9})[see Appendix~\ref{appendixB}]. 

In Step (ii) we achieve the conditional operation on the target atom depending on the state of the control atom. The Hamiltonian of the target atom is given by
\begin{eqnarray}\label{eq6}
H_t(t)&=&\frac{\Omega_{t0}(t)}{2}|0\rangle_t\langle r|+\frac{\Omega_{t1}(t)}{2}|1\rangle_t\langle r|+{\rm H.c.}\cr\cr
&=&\frac{\Omega_t(t)}{2}(\sin\frac{\Theta}{2}e^{i\varphi(t)}|0\rangle+\cos\frac{\Theta}{2}|1\rangle)_t\langle r| + {\rm H.c.},~~~~
\end{eqnarray}
in which $\Omega_{t0}(t)=|\Omega_{t0}(t)|e^{i\varphi_{t0}(t)}$,
$\Omega_{t1}(t)=|\Omega_{t1}(t)|e^{i\varphi_{t1}(t)}$,  $\varphi_{2}(t)=\varphi_{t1}(t)$,~$\varphi_{t0}(t)=\varphi(t)+\varphi_2(t)$,~$\Omega_t(t)=|\Omega_t(t)|e^{i\varphi_2(t)}$ with $|\Omega_t(t)|=\sqrt{|\Omega_{t0}(t)|^2+|\Omega_{t1}(t)|^2}$,~$\tan(\Theta/2)=|\Omega_{t0}|/|\Omega_{t1}|$. If $\Theta$ is time-independent, the Hamiltonian for target atom can be rewritten in the basis $ \{|b\rangle_t\equiv\sin(\Theta/2)e^{i\varphi}|0\rangle+\cos(\Theta/2)|1\rangle,  |d\rangle_t\equiv\cos(\Theta/2)|0\rangle-\sin(\Theta/2)e^{-i\varphi}|1\rangle,~|r\rangle_t\}$ as
\begin{eqnarray}\label{e13}
H_t(t)= \frac{\Omega_t}{2}|b\rangle_t\langle r| + {\rm H.c.},
\end{eqnarray}
where $|d\rangle$ is the dark state of the system that is decoupled from the dynamics.

We now consider the first case of step~(ii). If the control atom is initially in $|0\rangle$ state, it would not be excited after step~(i). As such, there is no RRI involved in the dynamics. Then, the evolution of target atom is controlled by Eq.~(\ref{e13}), which has the similar form to that of $H_c(t)$. Thus, one can use the similar method mentioned in step (i) to design the desired super-robust geometric operations in the subspace $\{|b\rangle_t, |r\rangle_t\}$. Specifically, we choose the time-dependent states as
\begin{eqnarray}\label{e14}
|\phi_{t1}(t)\rangle&\equiv& e^{i\Upsilon }\left[\cos\left(\frac{\theta_{t}}{2}\right)|b\rangle_t -\sin\left(\frac{\theta_{t}}{2}\right)e^{-i\eta_{t}}|r\rangle\right] ,
\cr\cr|\phi_{t2}(t)\rangle&\equiv& e^{-i\Upsilon }\left[\sin\left(\frac{\theta_{t}}{2}\right)e^{i\eta_{t}}|b\rangle_t +\cos\left(\frac{\theta_{t}}{2}\right)|r\rangle\right],~~~~
\end{eqnarray}
where $\Upsilon$, $\theta_{t}$ and $\eta_{t}$ are the time-dependent parameters.
Similar to the process in step (i),  the parallel transport and super-robust condition for the control parameters of step (ii) are given by,
\begin{eqnarray}\label{ConTar}
    &&\int_{\tau_1}^{\tau_1+\tau_{2}} \frac{\dot{\theta}_{t}}{2} \exp \left(-i \int_{\tau_{1}}^{\tau_{1}+t} \frac{\dot{\eta}_{t}}{\cos \theta_{t}} d t^{\prime}\right) d t=0, \cr\cr && \dot{\eta}_{t} \sin \theta_{t}=0 \ .
\end{eqnarray}
And the time-dependent laser parameters are determined by,
\begin{equation}\label{LaserTar}
\left|\Omega_{t}(t)\right| =\dot{\theta}_{t}, \quad\varphi_{2}(t) =\eta_{t}-\frac{\pi}{2} \ .
\end{equation}
Consequently, the geometric evolution operator in the subspace $\{|b\rangle_t, |r\rangle_t\}$ is obtained as equation~(\ref{Gate1}), with
$\boldsymbol{n}=\left[\sin \theta_{t}(0) \cos \eta_{t}(0), -\sin \theta_{t}(0) \sin \eta_{t}(0), \cos \theta_{t}(0)\right]$.

In this step, we implement the geometric operation $|b\rangle\rightarrow e^{i\pi}|b\rangle$ with $\Upsilon(\tau)=\pi$, which is equivalent to $U=-|b\rangle_t\langle b| - |r\rangle_t\langle r|$ when the initial state is $|b\rangle_t$. If we consider the decoupled dark state, the evolution operator in this step would be $U_{ii} = -|b\rangle_t\langle b| - |r\rangle_t\langle r| + |d\rangle_t\langle d|$. To achieve this goal with the super-robust pulse, without loss of generality, we design the Hamiltonian parameters as shown in Table~\ref{t2}.

\begin{table*}[h!t]
\renewcommand\arraystretch{1.5}
\caption{One group of possible parameters of the laser amplitude and phase to implement the super-robust geometric quantum operations in step (ii). $\tau_2$ is the total time of step~(ii). $|\Omega_{t}(t)|=(8\pi/\tau_{2})\sin^2[2\pi(t-t')/(\tau_{2}/2)]$.}
\begin{tabular}{ ccccc }
 \hline
 \hline
  & $~t~\in$~[$\tau_1,~\tau_1+\frac{\tau_2}{4}$]~ & $~t~\in$~($\tau_1+\frac{\tau_2}{4},~\tau_1+\frac{\tau_2}{2}$]~ & $~t~\in$~($\tau_1+\frac{\tau_2}{2},~\tau_1+\frac{3\tau_2}{4}$]~& $~t~\in$~($\tau_1+\frac{3\tau_2}{4},~\tau_1+\tau_2$]~\\
 \hline
 $\varphi_2(t)$	& 0	& $\frac{3\pi}{2}$ & 0&$\frac{3\pi}{2}$\\
 \hline
 ~~$t'$~~  & $\tau_1$ & $\tau_1$	& $\tau_1+\tau_2/2$ & $\tau_1+\tau_2/2$\\
   \hline\hline
 \end{tabular}\label{t2}
\label{symbols}
 \end{table*}

After this step, the operation $U_t=-|b\rangle\langle b| + |d\rangle\langle d|$ is achieved in the computational subspace, which can be re-expressed as
\begin{eqnarray}
U_t=\left(
\begin{array}{cc}
\cos (\Theta ) & -e^{i \varphi } \sin (\Theta ) \\
-e^{-i \varphi } \sin (\Theta ) & -\cos (\Theta ) \\
\end{array}
\right)
\end{eqnarray}
in the basis $\{|0\rangle_t,~|1\rangle_t\}$. Thus, one can set different groups of parameters $\Theta$ and $\varphi$ to realize various operations on the target atom. Concretely, one can choose $\{\varphi,~\Theta \}=\{0,~0 \}$ for $\sigma_{z}^{(t)}$~($\sigma_z$ operation on the target atom) operation, $\{\varphi,~\Theta \}=\{0,~-\pi/2\}$ for NOT operation and $\{\varphi,~\Theta \}=\{0,~-\pi/4\}$ for Hadamard operation, respectively.

We now consider the second case of step~(ii), i.e., when the control atom lies in $|1\rangle$ state initially. In this case the control atom would be excited after step~(i). Then the dipole-dipole interaction Hamiltonian
\begin{equation}
H_d = V|r'\rangle_c\langle R'|\otimes|R\rangle_t\langle r|+{\rm H.c.}
\end{equation}
would also be involved in controlling the dynamics of the whole system. The total Hamiltonian of step (ii) in the two-atom basis can thus be rewritten as
\begin{equation}\label{eq10}
H_{ii} = \frac{\Omega_t(t)}{2}|R'b\rangle\langle R'r| + V|r'R\rangle\langle R'r| + {\rm h.c.},
\end{equation}
in which the state $|mn\rangle\equiv|m\rangle_c\otimes|n\rangle_t$ and this abbreviation style would be used throughout this work. Eq.~(\ref{eq10}) has one dark state
\begin{equation}
|d\rangle_2 =\frac{1}{\mathcal{N}} \left(V|R'b\rangle - \frac{\Omega_{t}^*}{2}|r'R\rangle\right),
\end{equation}
where $\mathcal{N}$ is the normalized parameter of the dark state, and two bright states
\begin{eqnarray}
|b_\pm\rangle = \frac{1}{\sqrt{2}\mathcal{N}}\left(\frac{\Omega_t}{2}|R'b\rangle \pm\mathcal{N}|R'r\rangle+ V|r'R\rangle\right)~
\end{eqnarray}
with eigenvalues 0, $\pm \mathcal{N}$, respectively. Here $\mathcal{N}=\sqrt{V^2+\Omega_t^2/4}$ is the normalized parameter.

In principle, when the initial state is $|R'b\rangle$, the system would evolve along the dark state $|d_2\rangle$ and the population of the state $|r'R\rangle$ would increase when $\Omega_t$ increases slowly on the premise of meeting the adiabatic condition~[see Appendix \ref{appendixC}]. When $\Omega_t$ is set to be zero initially and finally, one can argue that $|R'b\rangle$ is still be populated when the adiabatic process is finished. We now analyze the phases accumulated in this process. The phase accumulated on the dark state can be classified as the dynamical phase $\varphi_{dy}$ and the geometric phase $\varphi_{ge}$, respectively, given by [see Appendix \ref{appendixD}]
\begin{equation}\label{eq13}
\varphi_{dy}=\int_{\tau_1}^{\tau_1+\tau_2}~_{2}\langle d|H_{ii}|d\rangle_2dt=0
\end{equation}
  and
\begin{equation}\label{eq14}
\varphi_{ge}=i\int_{\tau_1}^{\tau_1+\tau_2}~_{2}\langle d|\frac{\partial}{\partial t}|d\rangle_2dt=\int_{\tau_1}^{\tau_1+\tau_2} \frac{\Omega_{t}^2 \dot{\varphi}_{2}}{\Omega_{t}^2+4 V^2}dt=0.
\end{equation}
That is, the initial state $|R'b\rangle$ would keep invariant and accumulate no phase, thus the identity matrix $I_{t}$ is achieved for the evolution operator. To derive Eq.~(\ref{eq14}), we have supposed that $V$ is constant. However, in practical case, $V$ is determined by $C_3$ and $d$~($V=C_3/d^3$), where $C_3$ is the coefficient relevant to atom and Rydberg states, and $d$ is inter-atomic distance that linear in time. That is, the value of \emph{V} varies over time. In Appendix ~\ref{appendixE}, we show clearly that, the geometric phase is still zero in this case. One should note that although this case has the similar effect to the conventional Rydberg blockade, i.e., when control atom is excited, the state of the target atom would be invariant, the physical regime is completely different and the performance is better than that of blockade regime since the current scheme utilizes the adiabatic process that has less error if the adiabatic condition is satisfied well.

Step~(iii) is the reverse operation of step~(i). After these three steps, one acquires the operation
\begin{equation}\label{TWOQ}
    U = |0\rangle_c\langle0|\otimes U_{t} + |1\rangle_c\langle1|\otimes I_{t}.
\end{equation}
In general, Eq. (\ref{TWOQ}) is a nontrivial two-qubit gate. One can choose different parameters to realize the  CZ and CNOT gates, respectively. 

\section{Results and Discussions}\label{sec3}
In this section, we demonstrate through numerical results that the current scheme has stronger robustness to the Rabi frequency error and is also resilient to blockade error under the consideration of atomic spontaneous emission in contrast to the conventional blockade scheme. Moreover it has stronger robustness to the Rabi frequency error in contrast to the dark state schemes.

\subsection{Gate performance}

We take advantage of the Lindblad master equation to numerically simulate the performance of the scheme under decoherence, which can be written as
\begin{equation}
	\dot{\rho}=-i\left[H,~\rho \right]+\sum_{i=1}^{12}\Gamma_i\left(A_i\rho A_{i}^{\dagger}-\frac{1}{2}\{A_{i}^{\dagger}A_{i}\rho \} \right),
\end{equation}
where $\Gamma_i$ denotes decay or dephasing rate relevant to the dissipation process described by operator $A_{i}$. In our scheme $A_1 = |0\rangle_c\langle R'|$, $A_2 = |1\rangle_c\langle R'|$, $A_3 = |0\rangle_c\langle r'|$, $A_4 = |1\rangle_c\langle r'|$, $A_5 = |0\rangle_t\langle R|$, $A_6 = |1\rangle_t\langle R|$, $A_7 = |0\rangle_t\langle r|$ and $A_8 = |1\rangle_t\langle r|$ denote the decay processes of the excited states. $A_9=|R'\rangle_c\langle R'| - |0\rangle_c\langle0| - |1\rangle_c\langle1|$, $A_{10}=|r'\rangle_c\langle r'| - |0\rangle_c\langle0| - |1\rangle_c\langle1|$, $A_{11}=|R\rangle_t\langle R| - |0\rangle_t\langle0| - |1\rangle_t\langle1|$ and $A_{12}=|r\rangle_t\langle r| - |0\rangle_t\langle0| - |1\rangle_t\langle1|$ denote the dephasing processes.
	\begin{figure}[tbp]
		\centering
		\includegraphics[width=\linewidth]{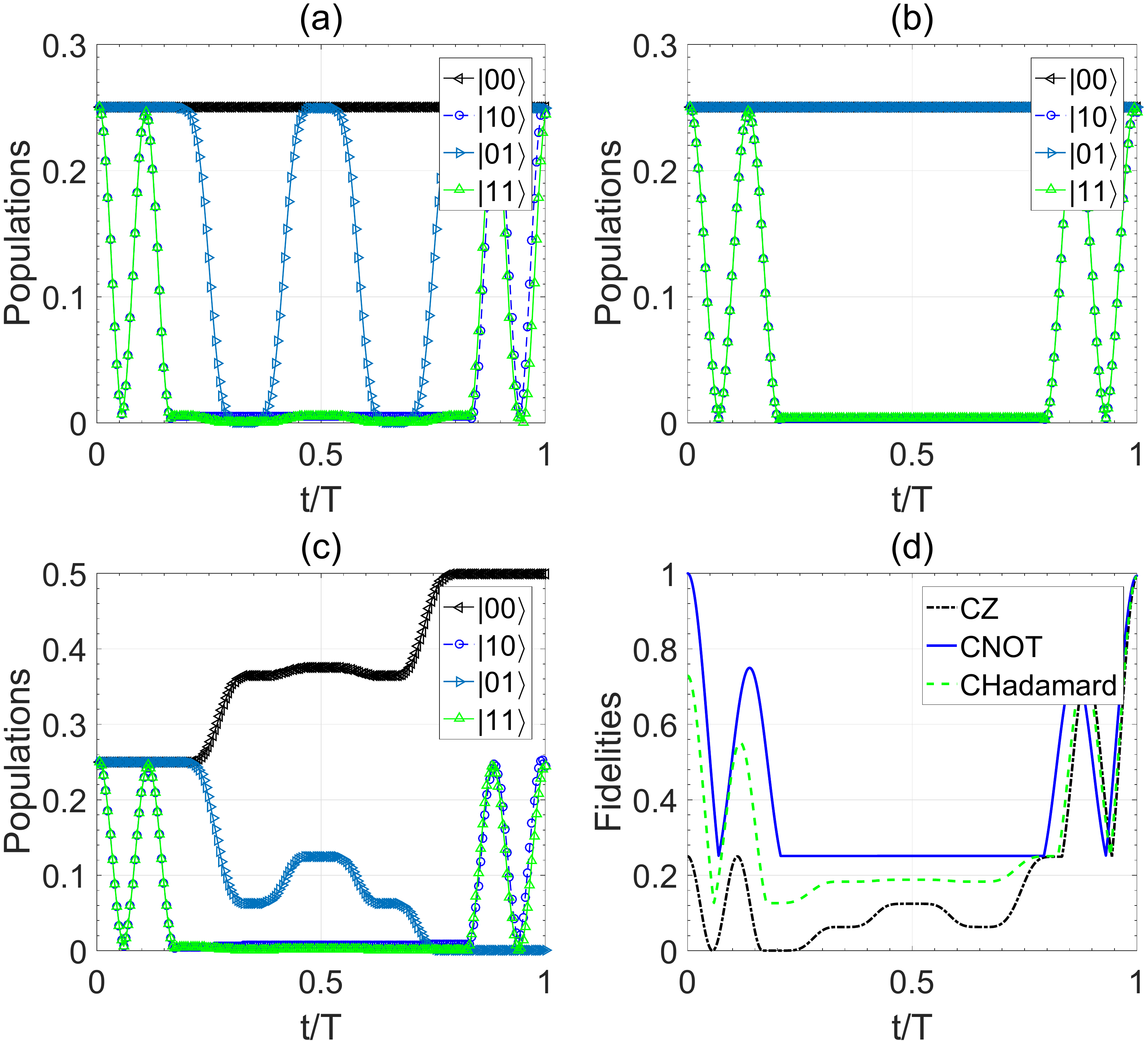}
		\caption{Population and fidelity of the constructed logic gates with the initial state $|\psi\rangle=(|00\rangle+|01\rangle+|10\rangle+|11\rangle)/2$. (a)[(b),~(c)] Populations of the CZ(CNOT, CHadamard) gate. (d) The fidelity of the constructed gates. The parameters are chosen as follows. $\Omega=2\pi\times8$~MHz, $|\Omega_{c}(t)|=\Omega$, $V=2\pi\times298$~MHz when the interatomic distance is set as $6~\mu$m for the chosen Rydberg levels. $\varphi_{c}(t)$ is shown in Eq.~(\ref{e10}). And $\varphi_{t1}(t)=\varphi_{2}(t)$ is shown in Eq.~(\ref{LaserTar}), $\varphi_{t0}(t) = \varphi_{t1}(t) + \varphi(t)$. $\Omega_{t0} = |\Omega_t(t)|\sin(\Theta/2)$, $\Omega_{t1} = |\Omega_t(t)|\cos(\Theta/2)$. $\varphi(t)$ and $\Theta$ are parameters determined by the concrete quantum logic gate we want to construct. $\Gamma_i$ is shown in the main text. For CZ, CNOT and CHadamard gate, ${\rm Max}[|\Omega_{t1}(t)|]$ is set as $2\Omega/3$.}\label{untitled}
	\end{figure}
    \begin{figure*}[htbp!]
		\centering
		\includegraphics[width=18cm]{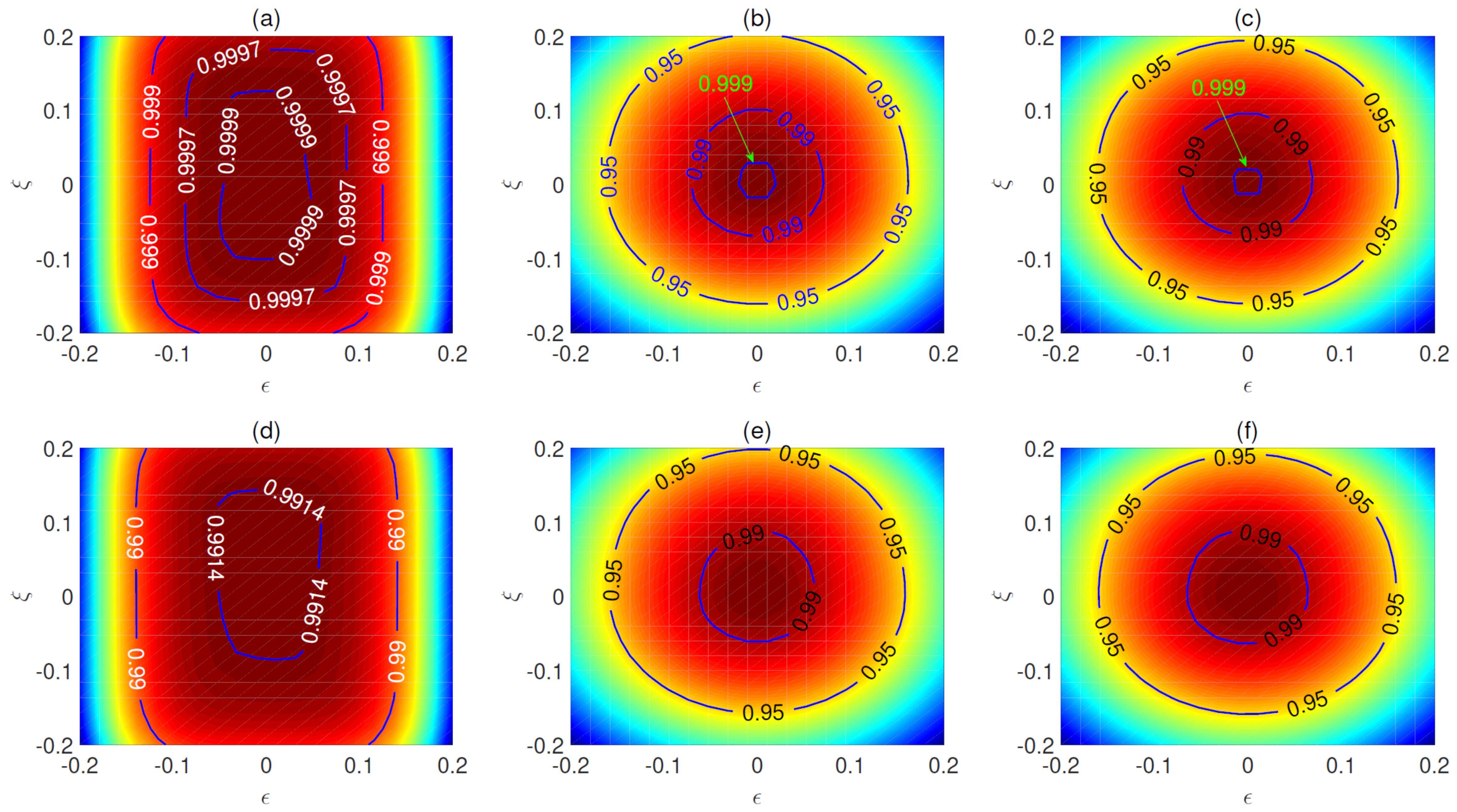}
		\caption{Fidelity of the CZ gate based on the current super-robust scheme(a)[(d)], the dark state scheme~\cite{klausdarkstate2017}(b)[(e)], and the conventional blockade scheme~\cite{DJP:2000}(c)[(f)], respectively, without (with) the consideration of dissipation. In panel~(a)[(d)], the parameters are chosen the same as Fig.~\ref{untitled}(a) with $\Gamma_i=0$(or not). In panel (b)[(e)], $|\Omega_{t}(t)|=(4\pi/\tau_{2})\sin^2[2\pi(t-\tau_1)/(\tau_{2}/2)](\tau_1<t\leqslant \tau_1+\tau_{2})$, and the scheme is described in Ref.~\cite{klausdarkstate2017}. In panel (c)[(f)], $\Omega_t = \Omega$, and the corresponding blockade scheme can be found in Refs.~\cite{DJP:2000,mtk2010}. In panels (a), (b) and (c), $\Gamma_i=0$. In panels (d) and (e), $\Gamma_1=\Gamma_2=2\pi\times0.425$~kHz, $\Gamma_3=\Gamma_4=2\pi\times0.213$~kHz, $\Gamma_5=\Gamma_6=2\pi\times0.169$~kHz, $\Gamma_7=\Gamma_8=2\pi\times0.336$~kHz, $\Gamma_9=\Gamma_{10}=\Gamma_{11}=\Gamma_{12}=2\pi\times1$~kHz. In panel (f), since the blockade scheme is three-energy-level structure, we consider $|R'\rangle$ for control atom and $|r\rangle$ for target atom, respectively. Thus, $\Gamma_3=\Gamma_4=\Gamma_5=\Gamma_6=\Gamma_{10}=\Gamma_{11}=0$, and the remaining rates are the same as panel (d).}\label{fig2}
	\end{figure*}
	In Fig.~\ref{untitled}, we plot the populations and fidelities of the constructed gates with a specific group of initial states. For the chosen energy level, the decoherence parameters are set as $\Gamma_1=\Gamma_2=2\pi\times0.425$~kHz, $\Gamma_3=\Gamma_4=2\pi\times0.213$~kHz, $\Gamma_5=\Gamma_6=2\pi\times0.169$~kHz, $\Gamma_7=\Gamma_8=2\pi\times0.336$~kHz~\cite{Beterov2009}. For the dephasing rate, we here temporarily set $\Gamma_9=\Gamma_{10}=\Gamma_{11}=\Gamma_{12}=2\pi\times1$~kHz at 0~K and using the relationship $n\sim\tau^3$ for the evaluation. The inter-atomic distance is set as 6 $\mu$m, which induces the RRI strength $V=2\pi\times298$~MHz for the chosen Rydberg states as considered in the caption of Fig.~\ref{fig1}. The results indicate that the final population and final fidelity agree well with the dynamics governed by the constructed gates.

\subsection{Robustness against Rabi control errors}

When constructing quantum logic gates theoretically, we often assume that the Rabi frequency is constant. However, in practice, there are some errors in the Rabi frequency due to the fluctuation of the laser intensity. From this point of view, in order to better demonstrate the super-robustness of the gate, we assume that the Hamiltonian would be written as follows. In steps~(i) and (iii), the Hamiltonian would be $H_c^{\rm error}=(1+\xi)H_c(t)$, and in step~(ii) the Hamiltonian is $H_t^{\rm error}=(1+\epsilon)H_t(t)$, where $\xi$ and $\epsilon$ are parameters regarding the Rabi frequency errors.
In the following, we use the average fidelity~\cite{White:07,NIELSEN2002249}
\begin{equation}
    F(\Xi,U)=\frac{\sum_j{\rm tr}\left(UU_{j}^{\dag}U^{\dag}\Xi(U_{j})\right)+d^2}{d^2(d+1)}
\end{equation}
to evaluate the performance of the present scheme, where $U_j$ is the tensor of Pauli matrices $II,~I\sigma_x,~\cdots~\sigma_z\sigma_z$, $U$ is the perfect phase gate, $d = 4$ for a two-qubit gate, and $\Xi$ is the trace-preserving quantum operation obtained through solving the master equation.

In Fig.~\ref{fig2}(a), (b) and (c), we plot the average fidelity of the current scheme,  the dark state scheme~\cite{klausdarkstate2017} and the conventional blockade scheme~\cite{DJP:2000} to demonstrate their robustness to both Rabi frequency errors $\xi$ and $\epsilon$ being $[0,~20\%]$. It can be seen that, as discussed in Ref.~\cite{klausdarkstate2017}, although the infidelity of the dark state scheme without Rabi frequency error can be $10^{-5}$ or even smaller, the robustness of the dark state scheme is not better than the current one when the Rabi frequency errors exist. The results indicate the super robustness feature of the current scheme in contrast to the other two schemes.

\begin{figure}[tbp]
	\centering
	\includegraphics[width=\linewidth]{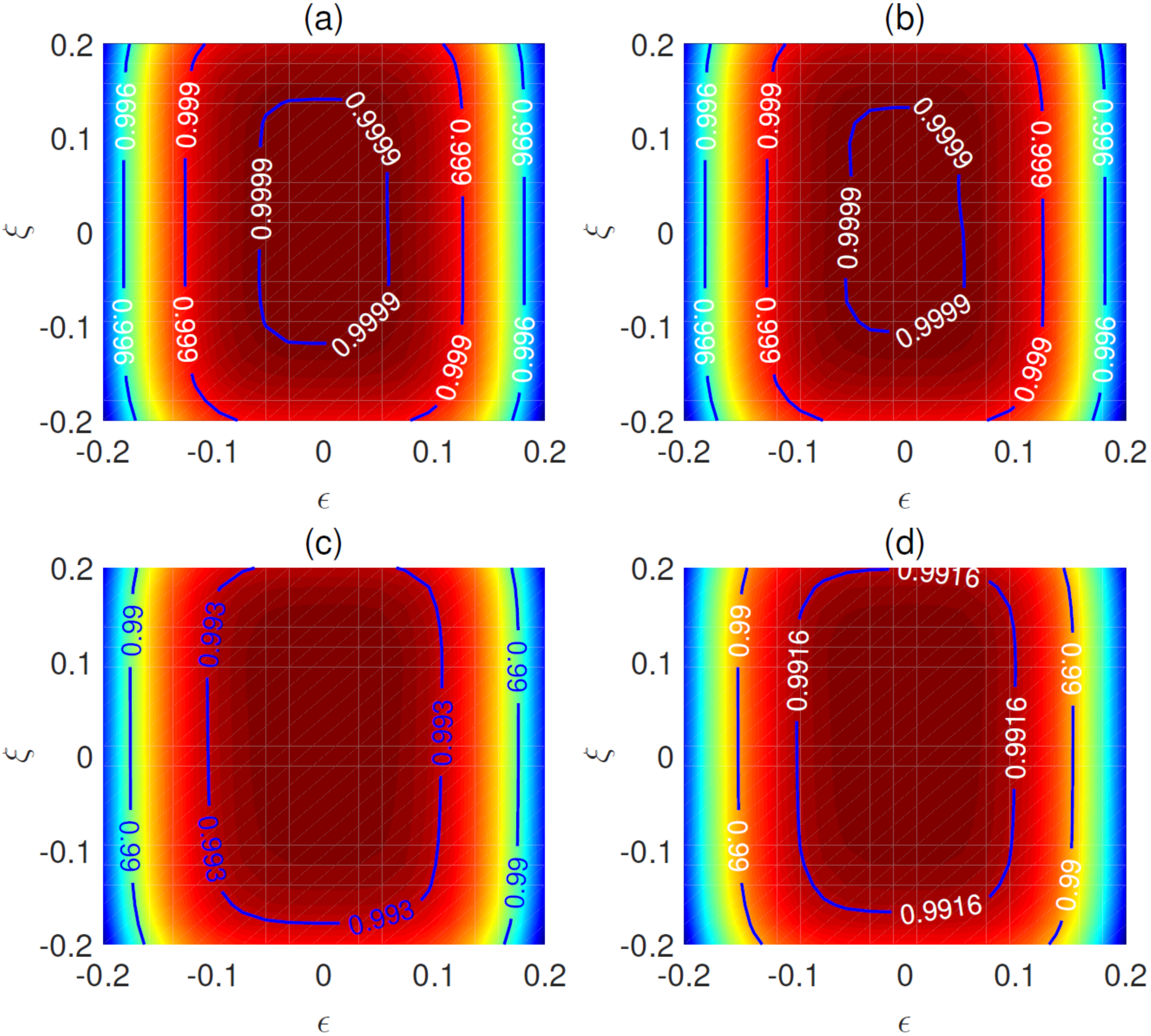} \caption{Fidelity of the CNOT~(a)[(c)] and CHadamard~(b)[(d)] gates based on the current super-robust scheme without~(with)~consideration of dissipation. The parameters are chosen the same as that of Fig.~\ref{untitled}(b) and (c), respectively, except for panels (a) and (b), $\Gamma_i$ is set as zero.}\label{fig3}
\end{figure}

To consider the decay and dephasing processes we choose the Rabi frequency, and the decay and dephasing rates the same as that in Fig.~\ref{untitled}. The inter-atomic distance is assumed to be 6 $\mu$m, which induces the RRI strength $V=2\pi\times298$~MHz for the chosen Rydberg states as considered in the caption of Fig.~\ref{fig1}. With these experimental parameters, the fidelity of the current scheme, the dark state scheme~\cite{klausdarkstate2017}, and the conventional scheme~\cite{DJP:2000} are plotted in Fig.~\ref{fig2}(d), (e) and (f), respectively. The results show that without Rabi frequency errors, the average fidelity of our scheme does not have an advantage, i.e., the average fidelity of the dark state, the current and the conventional blockade scheme are 0.9975, 0.9915, and 0.9982, respectively. That is due to the fact that our schemes require longer evolution time which enhances the influences of dissipation.  
However, when the Rabi error of each step is close to 20\%, our fidelity can still reach 0.99, and the fidelities of the other two schemes are just close to 0.95. 
In addition, we also plot the average fidelities of the CNOT and CHadamard gates in Fig.~\ref{fig3}, respectively, which also indicate the robustness of our scheme under the Rabi frequency errors and dissipation.
It should be noted that we have not considered the motion-induced dephasing here, which would no doubt decrease the average fidelity as shown in the following experimental considerations.

We now compare our scheme with the works presented in Refs.~\cite{wuprapplied2021,sunprapplied2021}. To obtain the Hamiltonian dynamics, Refs.~\cite{wuprapplied2021,sunprapplied2021} utilize second-order perturbation theory twice during the derivation of the effective Hamiltonian. The current scheme does not utilize the perturbation theory for the Hamiltonian, which means the dynamics could be faster. For the operation steps, the schemes in Ref.~\cite{wuprapplied2021,sunprapplied2021} require one step while the current one requires three steps. For the optimal geometric quantum computation method, Ref.~\cite{wuprapplied2021} employs the zero-systematic error method~\cite{Ruschhaupt_2012} while Ref.~\cite{sunprapplied2021} considered the time-optimal technology~\cite{timeoptimal2015}. In our scheme, the super-robust pulse that can limit error to fourth-order is utilized. Ref.~\cite{wuprapplied2021} aims to implement multiple-qubit gate, and Ref.~\cite{sunprapplied2021} constructed three-qubit gate, while we construct the robust two-qubit gate.

\subsection{Blocked-error resilience analysis}
In the conventional blockade scheme, the fidelity of the constructed gate would decrease when the RRI strength is not strong enough, leading to the blockade error. This blockade error is proportional to the square of $\Omega/V$~\cite{shi2017free,saffmanblockadeerror}, where $\Omega$ is the Rabi frequency and $V$ the RRI strength. Thus, when $V\sim\Omega$, the blockade error is large enough that decreases qualities of the scheme.

\begin{figure}[tbp]
	\centering
	\includegraphics[width=\linewidth]{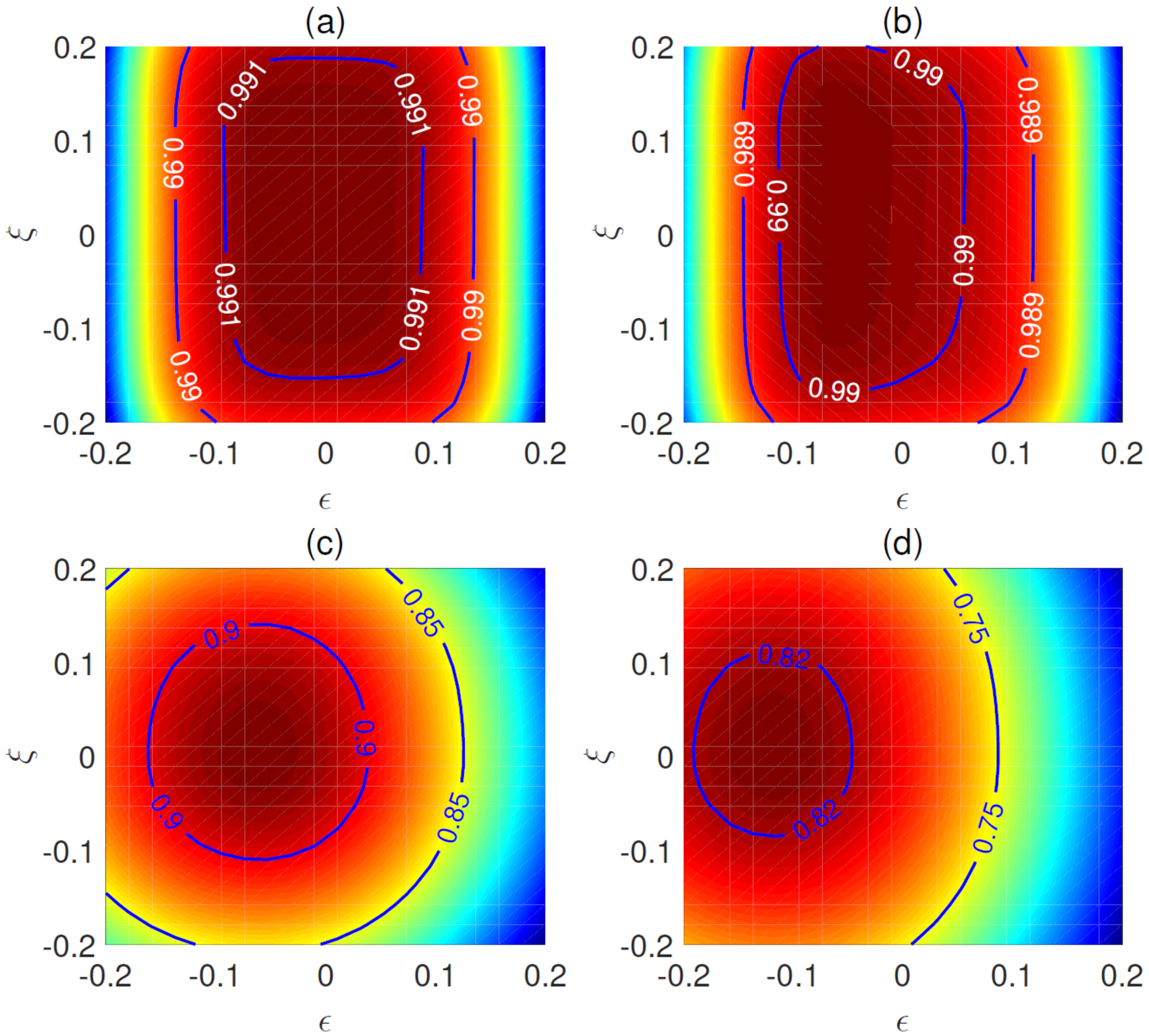}
	\caption{Fidelity of the super-robust CZ gate with $V\sim2\max[|\Omega|]$~(a) and $V\sim\max[|\Omega|]$~(b) under dissipation. Fidelity of the conventional blockade scheme with $V\sim2\max[|\Omega|]$~(c) and $V\sim\max[|\Omega|]$~(d) under the consideration of dissipation. The rest parameters are chosen the same as that of Fig.~\ref{fig2}.}\label{figblockadeerror}
\end{figure}

In Fig.~\ref{figblockadeerror}, we show the average fidelity of the CZ gate with weak RRI strength. One can see that, for the conventional blockade scheme, the blockade error significantly influences the performance. While for our scheme, the average fidelity is still very high even with weak RRI strength and large Rabi frequency errors, implying the robustness of the scheme to the blockade errors as well as the Rabi frequency errors.

\section{Experimental considerations}\label{sec4}

\subsection{Excitation process and concrete laser parameters}
\begin{figure}[tbp]
	\centering
	\includegraphics[width=6cm]{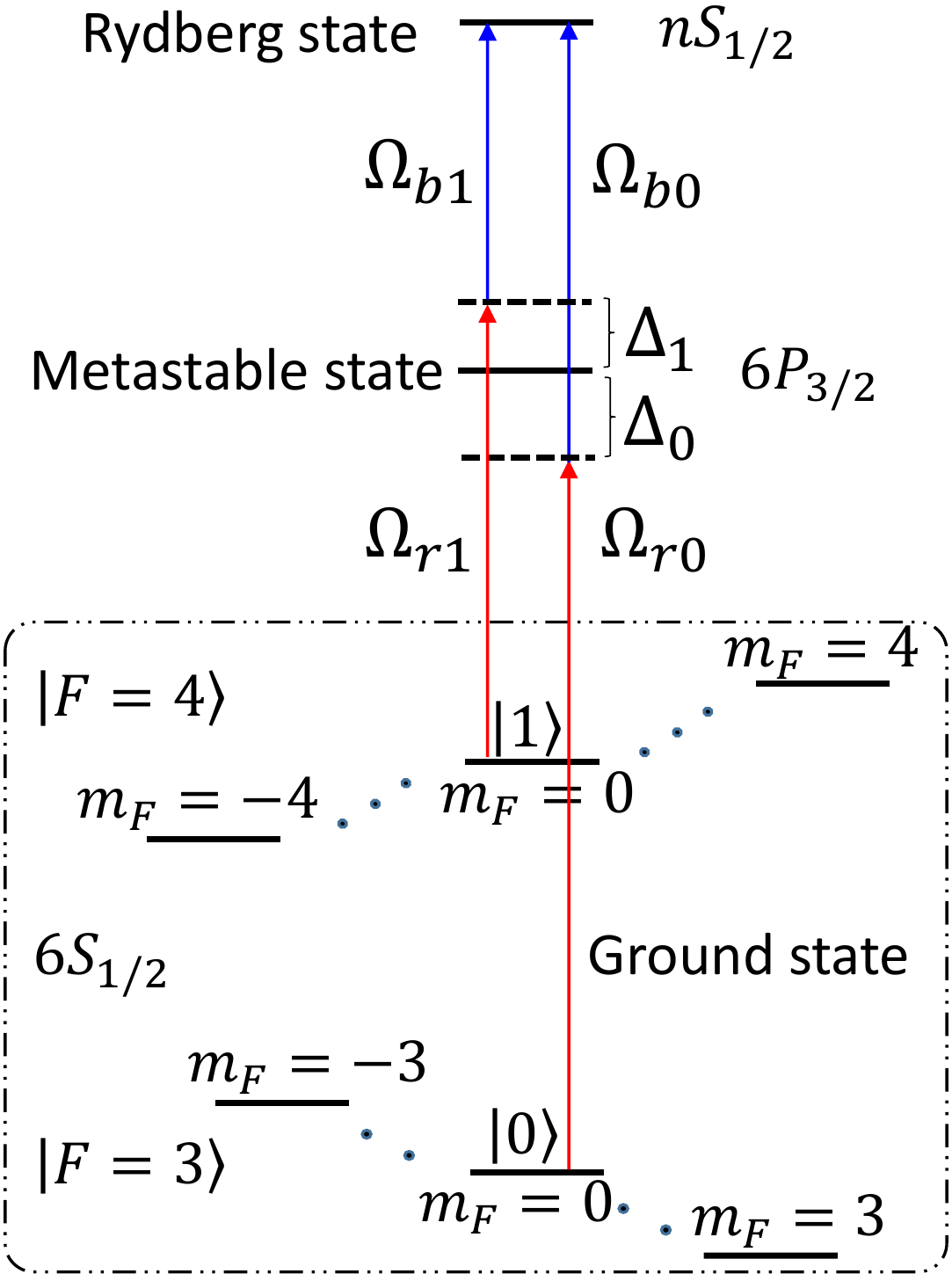}
	\caption{Energy levels for the considered $^{133}$Cs atom. $\Omega_{ri}$ and $\Omega_{bi}$ $(i=0,~1)$ denote the Rabi frequencies of the two-photon process $|i\rangle\rightarrow|ns_{1/2}\rangle$ with detuning $\Delta_i$, respectively. And $\Delta_i\gg\{\Omega_{ri}, \Omega_{ti} \}$ should be satisfied. For the control atom, the two-photon process from $|0\rangle$ to $|nS_{1/2}\rangle$ is not exist, and the parameters satisfy $\Omega_{c}=\Omega_{r1}\Omega_{b1}/(2\Delta_1)$. For the target atom, the parameters satisfy $\Omega_{ti}=\Omega_{ri}\Omega_{bi}/(2\Delta_i)$.}\label{fig6}
\end{figure}

The excitation to Rydberg state can be implemented by single-photon process~\cite{jau2016entangling,Hankin2016}. In this case the Rydberg state should be considered as $|np\rangle$ level due to the selection-rule. In this scheme, we consider the two-photon excitation process. As shown in Fig.~\ref{fig6}, the energy levels are chosen as $|0\rangle\equiv |f=3,~m_f=0\rangle$ and $|1\rangle\equiv |f=4,~m_f=0\rangle$, two long-lived ground states of Cs atom clock states. $|R'\rangle\equiv|101s_{1/2},~m=1/2\rangle$ and $|r'\rangle\equiv|101p_{3/2},~m=3/2\rangle$ are two Rydberg states of the control atom, and $|R\rangle\equiv|109p_{3/2},~m=3/2\rangle$ and $|r\rangle\equiv|109s_{1/2},~m=1/2\rangle$ are Rydberg states of the target atom. The resonant dipole-dipole interaction can be achieved with $C_3=64.4$~GHz$\cdot\mu m^3$ under the electric field $E=15.4$~V/m~\cite{klausdarkstate2017}. Alternatively, there are other choices of the Rydberg level for experiments. For instance, one can also choose $|R'\rangle\equiv|112s_{1/2},~m=1/2\rangle$, $|r'\rangle\equiv|111p_{3/2},~m=3/2\rangle$, $|R\rangle\equiv|101p_{3/2},~m=3/2\rangle$ and $|r\rangle\equiv|101s_{1/2},~m=1/2\rangle$. The resonant dipole-dipole interaction can be achieved with $C_3=65.3$~GHz$\cdot\mu m^3$ under the electric field $E=5.36$~V/m~\cite{klausdarkstate2017}.

The laser parameters are as follows. For the control atom, the wavelengths for $|1\rangle\rightarrow|P\rangle$ and $|P\rangle\rightarrow|R'\rangle$ are set as 852 and 509~nm, respectively. The Rabi frequencies are assumed to be $\Omega_{r1}=2\pi\times245$~MHz and $\Omega_{b1}=2\pi\times80$~MHz, respectively, and the detuning is 1.225~GHz. To achieve this goal, the power and waist are 1~$\mu$W and 3.6~$\mu$m for the red laser and 80~mW and 3~$\mu$m for the blue laser, respectively. For the target atom, the wavelengths are the same as those of the control atom, also for the $|0\rangle\rightarrow|r\rangle$ process. Rabi frequency is time-dependent, we here only consider how to achieve the maximal values, and the time-dependent characteristic can be achieved by tuning some of the laser parameters, such as the power. The Rabi frequencies are set as $\Omega_{r1}=2\pi\times245$~MHz and $\Omega_{b1}=2\pi\times80$~MHz, respectively. The detuning is set as 1.225~GHz. To achieve this goal, the power and waist are 1~$\mu$W and 3.6~$\mu$m for red laser and 80~mW and 2.7~$\mu$m for blue laser, respectively (for CZ gate, $\Omega_{r0}=\Omega_{b0}=0$ for the target atom). The inter-atomic distance is 6~$\mu$m.
\begin{figure}[tbp]
	\centering
	\includegraphics[width=\linewidth]{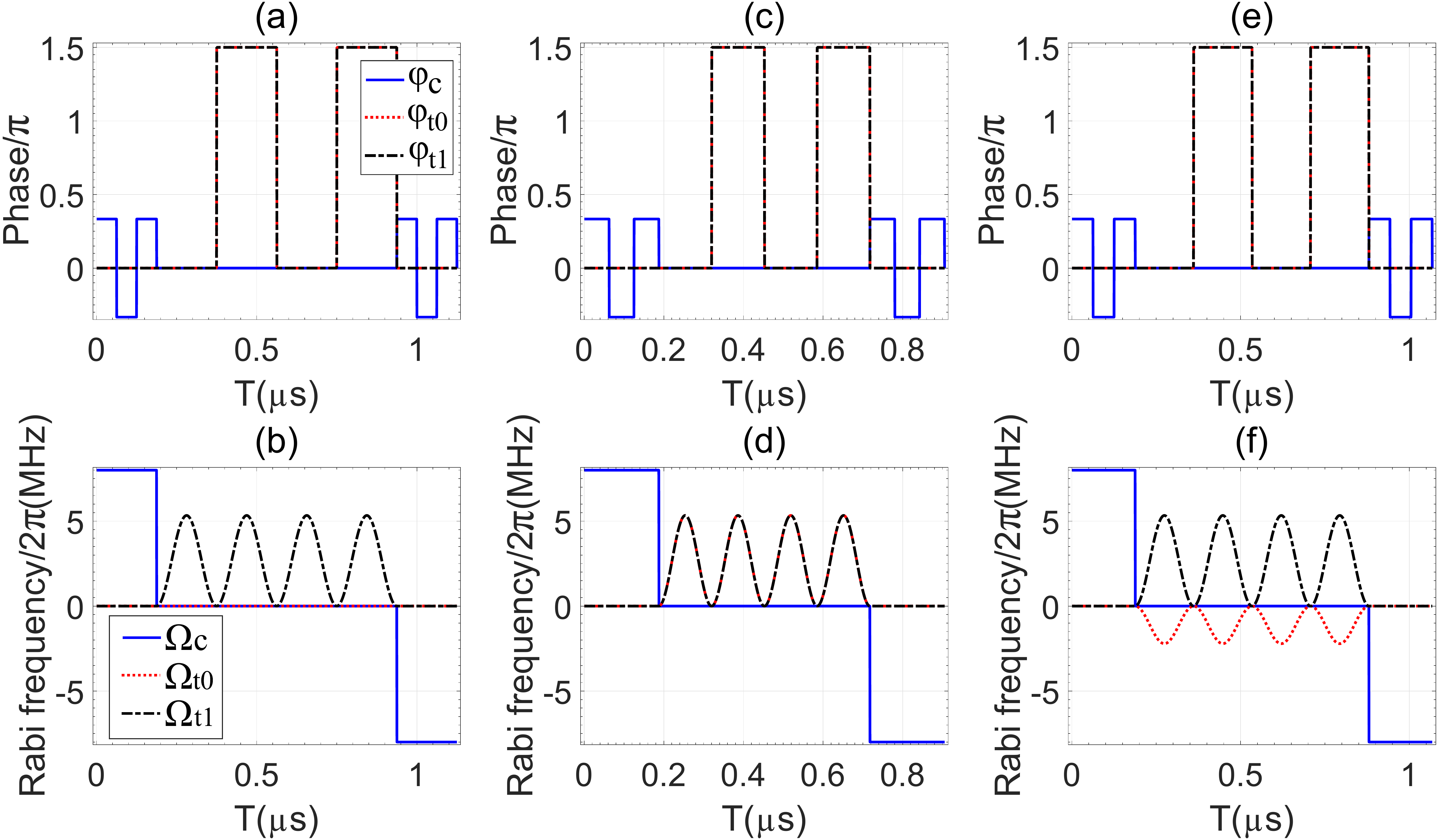}
	\caption{Shape of the Rabi frequency and the corresponding phase. (a)[(c) (e)] phase of the laser for CZ~(CNOT, CHadamard) gate. (b)[(d) (f)], Rabi frequency of the CZ~(CNOT, CHadamard) gate. It should be noted that the maximum value of the Rabi frequency of the target atom can be random, provided that the laser power and beam waist parameters are allowed experimentally and the adiabatic conditions are met.}\label{pulse}
\end{figure}
And the pulse shape of the laser used in this scheme is shown in Fig.~\ref{pulse}.

\subsection{Effectiveness of the two-photon excitation process and influence of larger Rabi errors}
For the excitation processes to Rydberg state, we considered here is the effective two-photon process. Thus, it is worthwhile to discuss the validity of the excitation process with the parameters used in this work.
\begin{figure}[tbp]
	\centering
	\includegraphics[width=0.85\linewidth,height=0.28\textwidth]{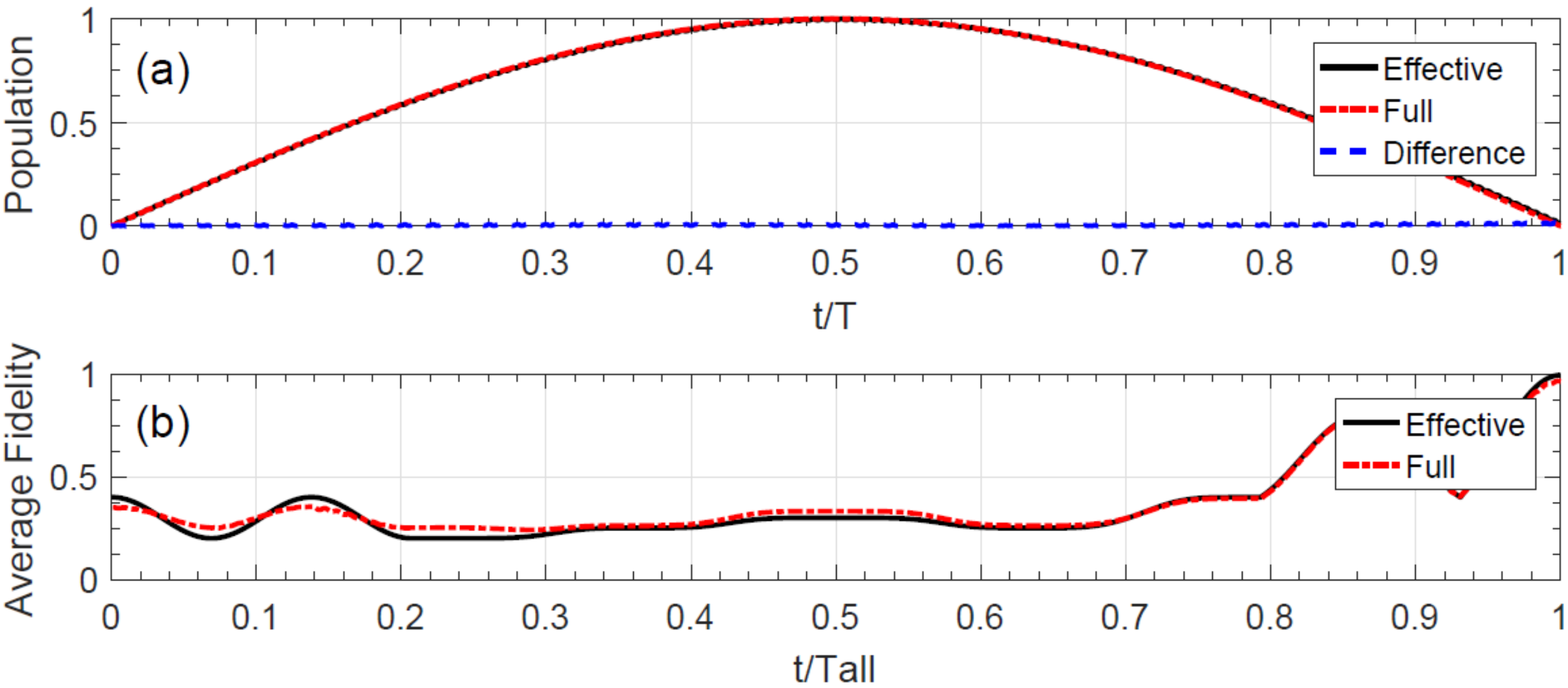}
	\caption{(a) Populations of excited state with respect to evolution time for the two-photon (full process) and effective processes, respectively, without consideration of dissipation. The dotted line denotes the difference of these two process. (b) Average fidelity of CNOT gate based on the effective and the full two-photon excitation process. The decay rates are considered the same as that in Fig.~\ref{fig3}. The decay of the intermediate state $|\mathcal{P}\rangle$ is considered as $2\pi\times3.2$~MHz.}\label{fig7}
\end{figure}
We here consider the two-photon excitation process $|g\rangle\xrightarrow[\Delta]{\Omega_{r}}|\mathcal{P}\rangle\xrightarrow[-\Delta]{\Omega_b}|\mathcal{R}\rangle$ (full process) and the effective process $|g\rangle\xrightarrow[\rm resonant]{\Omega_r\Omega_{b}/(2\Delta)}|\mathcal{R}\rangle$ (effective process). In Fig.~\ref{fig7}(a), we simulate the full and effective process with the parameters $(\Omega_{r},~\Omega_b,~\Delta)=2\pi\times(245~{\rm MHz},~80~{\rm MHz},~1.225~{\rm GHz})$, from which one can see that the full and effective process coincide with each other very well. In Fig.~\ref{fig7}(b), we plot the average fidelity of the proposed CNOT gate with the consideration of dissipation with full and effective processes, the result also demonstrates the validity of the effective model. 

\begin{figure}[tbp]
	\centering
	\includegraphics[width=\linewidth]{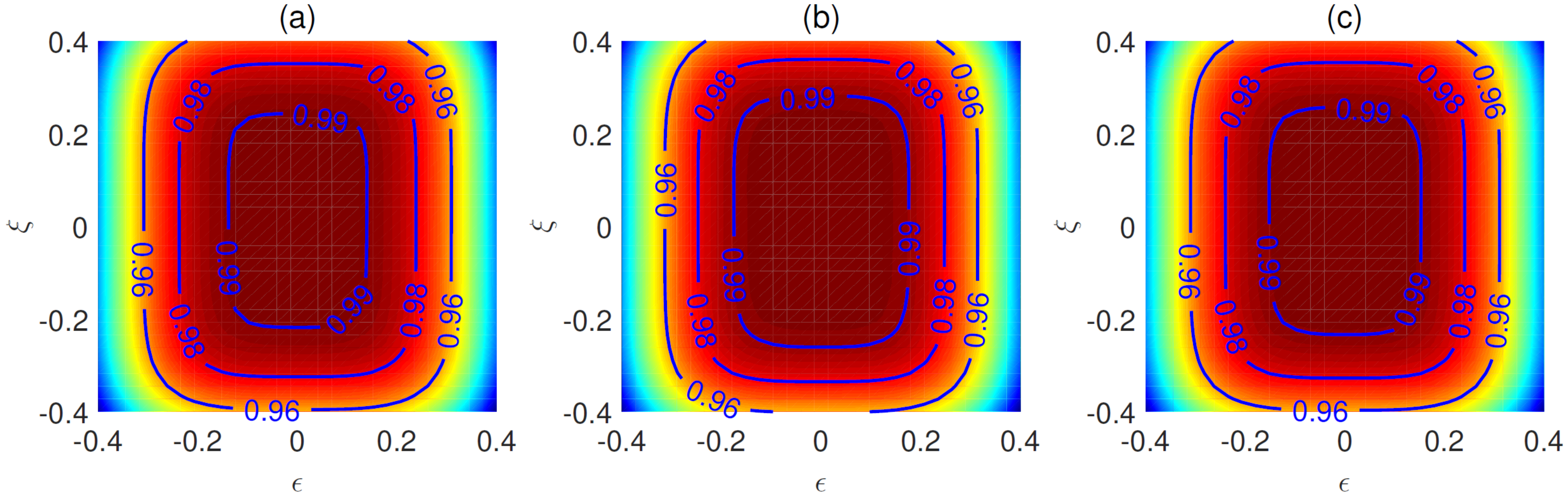}
	\caption{(a)[(b), (c)]Average fidelity of the proposed CZ(CNOT, CHadamard) gate with the maximal Rabi error 40\%. The parameters are the same as Fig.~\ref{fig2}(d)[Fig.~\ref{fig3}(c), Fig.~\ref{fig3}(d)].}\label{fig8}
\end{figure}

The large Rabi frequency of two-photon process requires the narrow waist of laser, which may change the Rabi error more than 20\% that we discussed in the main text. For this point, we plot the average fidelity of the constructed gates under dissipation with Rabi error as large as 40\% in Fig.~\ref{fig8}. The fidelity is above 96\% (without consideration of motion-induced dephasing) in most of the regions when both of $\epsilon$ and $\xi$ are as large as 40\%, which further proves the robustness of the scheme.

In fact, the single-photon excitation process to Rydberg state is also available for $^{133}$Cs atom in our scheme. It should be noted that one typically cannot access very high Rydberg state in this case. For instance, $n=64$ is achievable in experiment~\cite{jau2016entangling}.

\subsection{Leakage error to neighboring Rydberg states due to dipole-dipole interaction}\label{s4c}

From a practical point of view, we here consider the Rydberg state leakage error~\cite{saffmandrag}. The definition of this error is $P = 1-|\langle \Psi'|\psi(t)\rangle|^2$, where $|\Psi'\rangle = |R'\rangle_c\otimes|1\rangle_t$ is the initial state and $|\psi(t)\rangle$ is the system state after the evolution of Hamiltonian in step~(ii).  For our chosen Rydberg levels, the possible leakage channels contains $|101s_{1/2},~m=1/2; 109s_{1/2},~m=1/2\rangle\leftrightarrow|101p_{3/2},m = -1/2;109p_{1/2},m = -1/2\rangle$ and $|101p_{3/2},~m=3/2; 109p_{3/2},~m=3/2\rangle\leftrightarrow|99d_{5/2},m = 5/2;108d_{5/2},m = 5/2\rangle$~\cite{klausdarkstate2017}. In table~\ref{tab3}, we show the maximal and the average leakage errors when performing these three gates on the initial state, which shows the leakage error is negligible.

\begin{table}
	\caption{\label{tab3}Leakage error of Rydberg states. We here consider two leakage channels with two groups of energy-adjacent Rydberg states based on the spirit of Refs.~\cite{saffmandrag,klausdarkstate2017}. Leakage channel one is $|R'r\rangle\leftrightarrow|101p_{3/2},m = -1/2;109p_{1/2},m = -1/2\rangle$, the strength and detuning for this channel are $2\pi\times120$ and $2\pi\times65$~MHz, respectively. Leakage channel two is $|r'R\rangle\leftrightarrow|99d_{5/2},m = 5/2; 108d_{5/2},m = 5/2\rangle$, the strength and detuning for this channel are $2\pi\times156.8$ and $2\pi\times190$~MHz, respectively.}
	\begin{ruledtabular}
		\begin{tabular}{cccc}
						 &CZ\footnote{The results are achieved with the fourth-order Runge-Kutta method.} & CNOT\footnotemark[1] & CHadamard\footnotemark[1]\\ \hline
	  Maximal& $8.08\times10^{-5}$ & $8.11\times10^{-5}$ & $8.05\times10^{-5}$\\
	   Average & $3.01\times10^{-5}$ & $3.02\times10^{-5}$  & $3.01\times10^{-5}$ \\
			\end{tabular}
	\end{ruledtabular}
\end{table}

\subsection{Excitation error to the neighboring Rydberg states due to the imperfection excitation process}
\begin{figure}[tbp]
	\centering
	\includegraphics[width=\linewidth]{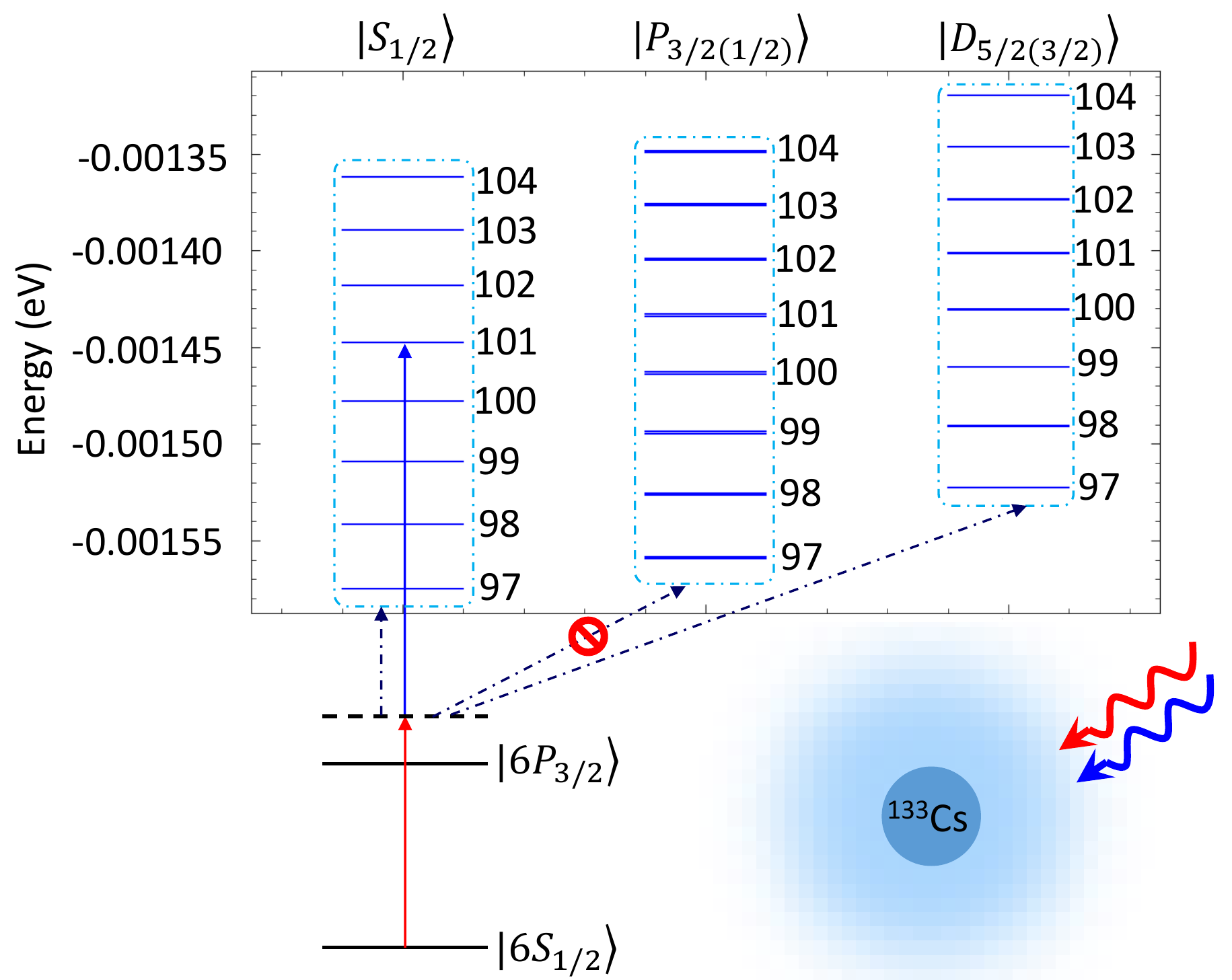}
	\caption{Main possible excitation leakage channels~(the states in the left and right rectangles enclosed by the dotted-dashed line are possible leakage states) for the control atom. The ideal excited state is $|101S_{1/2}\rangle$, and we consider the principle quantum number from 97 to 104. The relevant data is get by the package in Python~\cite{python}. And the similar process can also be plotted for the target atom.}\label{excitationerror}
\end{figure}
We now consider the leakage error to the neighboring Rydberg states. As shown in Fig.~\ref{excitationerror}, the principle quantum number from $n = 97$ to 104 is considered. The energy detuning $\delta^{\rm Leakage}_k=E^{\rm leakage~state}_k-E_{|101S_{1/2}\rangle}$ is the main influence factor to the excitation error, where $E^{\rm leakage~state}_k$ denotes the energy of the $k$-th leakage state in Fig.~\ref{excitationerror}. The leakage probability to the $k$-th state can be approximately described by $P_{\rm error} \approx \Omega^2/(\delta^{\rm Leakage}_k)^2$(see Appendix~\ref{af} for detail), here $\Omega$ denotes the effective two-photon Rabi frequency from ground state to the ideal Rydberg state. Thus, one can get the total excitation error as 
\begin{equation}
    P_{\rm error}^{\rm excitation} \approx x\times\sum_{k}(\frac{\boldsymbol d_{|6P\rangle\rightarrow k{\rm th}~{{\rm state}}}}{\boldsymbol d_{|6P\rangle\rightarrow |101(109)S_{1/2}\rangle}}\Omega)^2/(\delta^{\rm Leakage}_k)^2,
\end{equation}
where $x=3$~($x=2$) denotes the excitation times for control~(target) atom during the control-error-robust operations and $\boldsymbol d$ is the relevant transition dipole moment. After substituting the data in Fig.~\ref{excitationerror} and relevant dipole moment, one can get the sum of $P_{\rm error}^{\rm excitation}$ for control and target atoms is about $8\times10^{-4}$ when $\Omega=2\pi\times8$~MHz. And if one can consider all of the excitation leakage channels with larger detuning, the order of magnitude of the result can be conservatively estimated to be $10^{-3}$, which is larger than the leakages in Rydberg states due to dipole-dipole interactions discussed in Sec.~\ref{s4c}.

\subsection{The effects of motion-induced dephasing}

So far, we mainly focus on the influence of the Rabi control error and blockade error. However, as mentioned in Ref.~\cite{graham2019rydberg}, the dephasing error induced by motion when exciting the neutral atom from ground to Rydberg state is another factor that limits the fidelity and the geometric phase may not robust to this error~\cite{PRXQuantumDong}. The most accurate way to analyze the effects of such motion is to use quantum mechanical treatments~\cite{photonrecoil2021}. In this subsection, we will treat the motion of atom ballistic~\cite{shi2020motion} and propose to use the spin-echo to suppress the influence. This will provide some reference for the experiments. 

\begin{figure}[tbp]
	\centering
	\includegraphics[width=\linewidth]{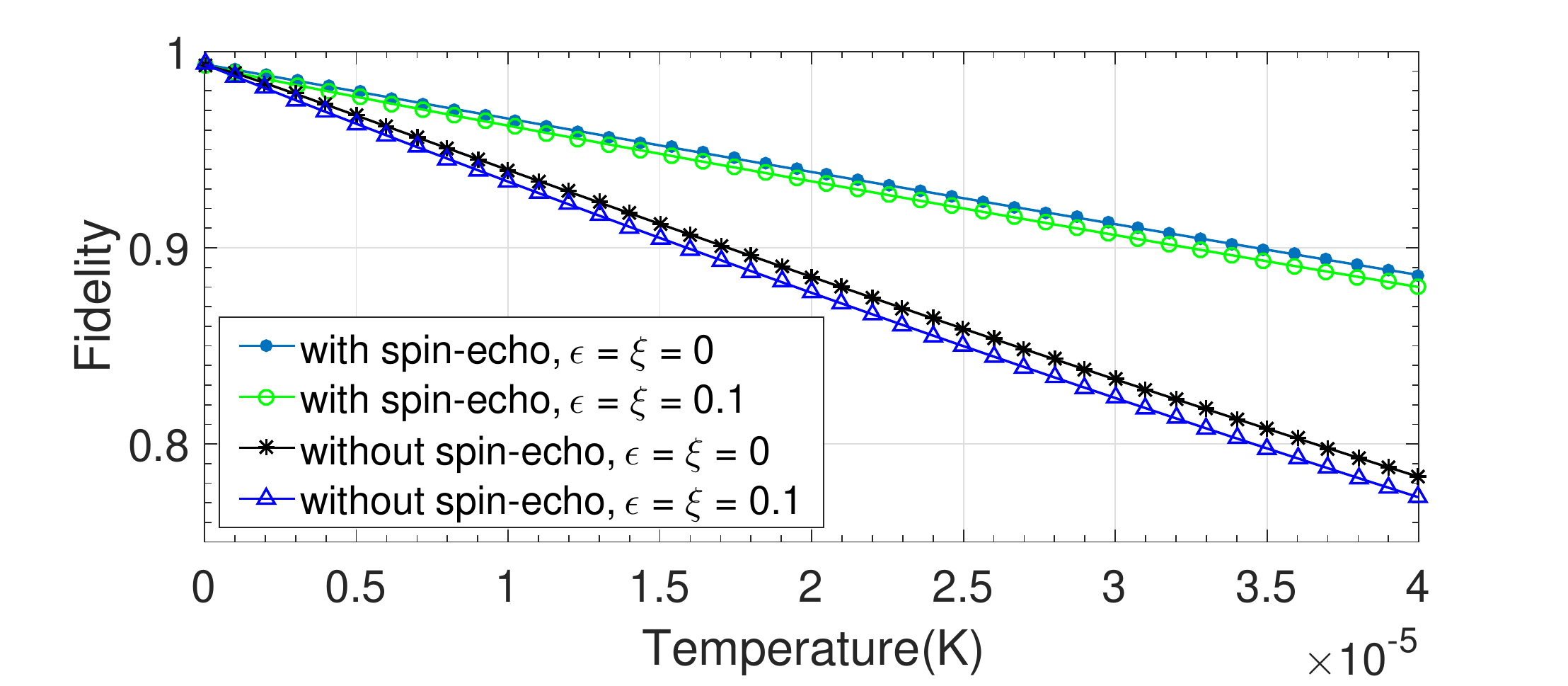}
	\caption{Average fidelity of the CNOT gate versus temperature with the consideration of dissipation. The parameters are the same as Fig.~\ref{fig3}(c).}\label{fig10}
\end{figure}

We take the control atom as an example to analyze the process, and a similar process applies to the target atom as well. Here for simplicity we do not consider the effect of the temperature on atomic spontaneous emission. The Rabi frequency should be modified as $\Omega_{c}e^{i\boldsymbol{kv}t}$ (for two-photon process with the intermediate state being large detuned, this can be calculated by the second-order perturbation theory) when we consider the motion of atoms, where $\boldsymbol{k}$ is the effective wave vector, $\boldsymbol{v}$ is the atomic velocity and can be approximately calculated as $|\boldsymbol{v}| = \sqrt{k_b{\mathcal T}/m}$ with $k_b$, $\mathcal{T}$ and $m$ being Boltzmann parameter, atomic temperature and mass, respectively. The atom initial position is not considered because one can set it as a relative position and thus has no influence for the process~\cite{shi2020motion}. The first strategy is the spin-echo method~\cite{saa2018}, for the control atom we change the Doppler detuning at the midpoint of the first step and the third step. While for the target atom we also change the Doppler detuning at the midpoint of the second step. This can be done be modulating the direction of the wave vector. In Fig.~\ref{fig10}, we take advantage of CNOT gate as an example to show the performance of the scheme under Doppler shift with the consideration of spin-echo technology, the result show that the scheme has a significant improvement with the consideration of spin-echo technology. Specifically, when the atomic temperature is at about $10~\mu$K($35~\mu$K), the average fidelity is improved from 0.93(0.8) to 0.97(0.9). We have to admit that this value of fidelity is slightly lower than that in Ref.~\cite{levine2019parallel}, due to the fact that our control-error-resilient pulse requires more evolution time. On the other hand, when the system error is around 10\%, our scheme is still able to maintain around this value, which is the main feature of our work.

\begin{figure}[tbp]
	\centering
	\includegraphics[width=\linewidth]{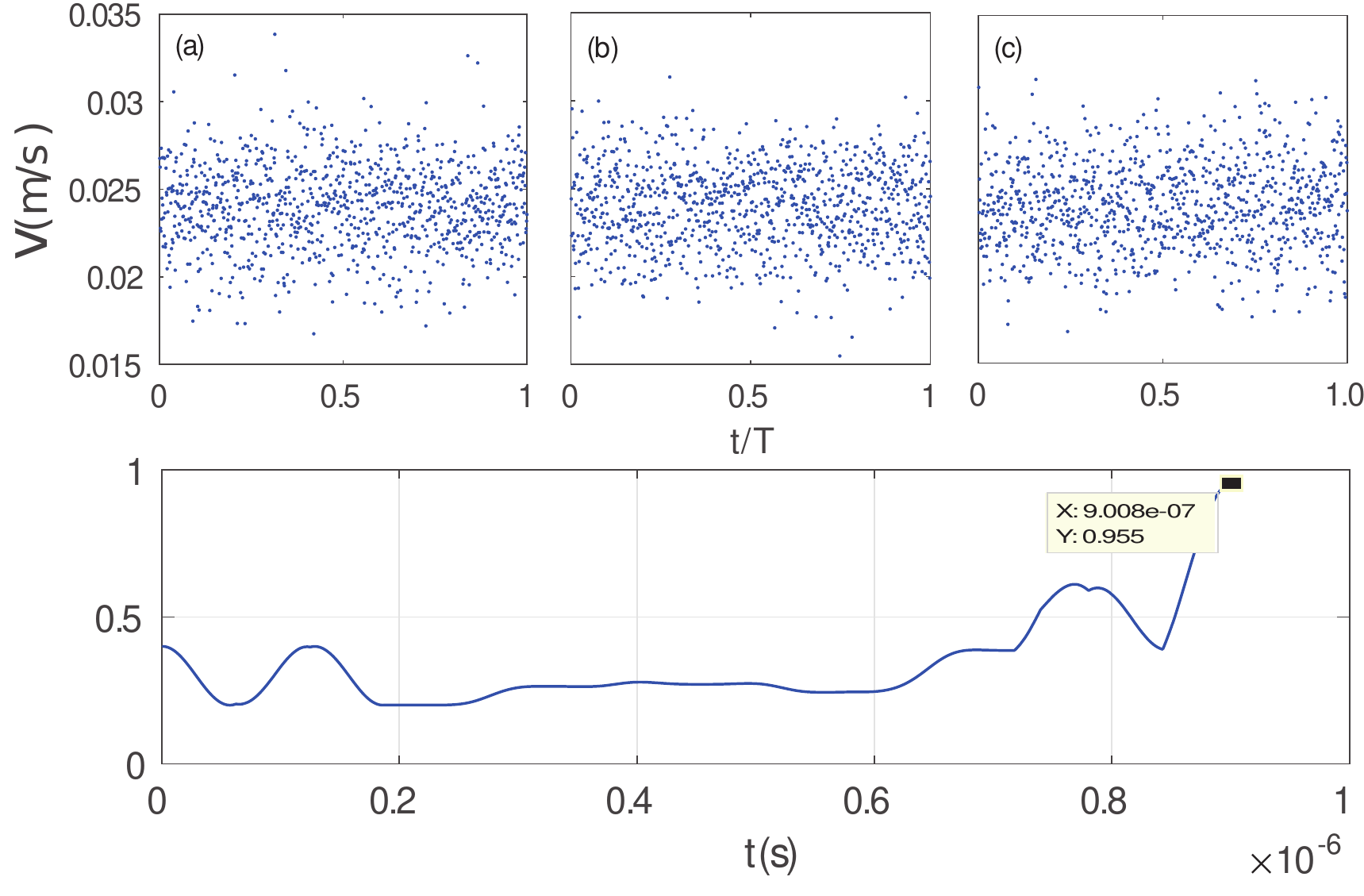}
	\caption{(a)[(c)]~Random speed of the control atom in step~(i)[(iii)]. (b)~Random speed of the target atom versus time in step~(ii). The atomic temperature is set as $\mathcal{T} = 10~\mu$K, the average velocity is calculated by $|\boldsymbol{v}|=\sqrt{k_b\mathcal{T}/m}$, and the variation is set as $0.1|\boldsymbol{v}|$. (d)~Average fidelity of the CNOT gate versus the evolution time with the consideration of spin-echo with the Rabi error~(control error) being 10\%. The rest parameters are the same as that in Fig.~\ref{fig3}(c).}\label{figrandom}
\end{figure}

We treat the speed as a constant in the above analysis, while in reality, the speed varies randomly within a certain range, which will undoubtedly reduce the fidelity of the scheme. In this case, we consider the Gauss distribution of the atomic velocity in each step individually when the temperature of the atoms is 10~$\mu$K, and use spin-echo technology in the process to numerically solve the master equation. The result in Fig.~\ref{figrandom} show that the average fidelity can be 0.955 even with the individual random velocities and the control error being 10\% under dissipation. We also make simulations with the control error being 20\% and random velocities, the results show that the average fidelity can be as 0.934. These results demonstrate the robustness of the scheme to control errors under realistic experimental conditions.

\subsection{Other practical considerations}
\begin{figure}[tbp]
	\centering
	\includegraphics[width=\linewidth]{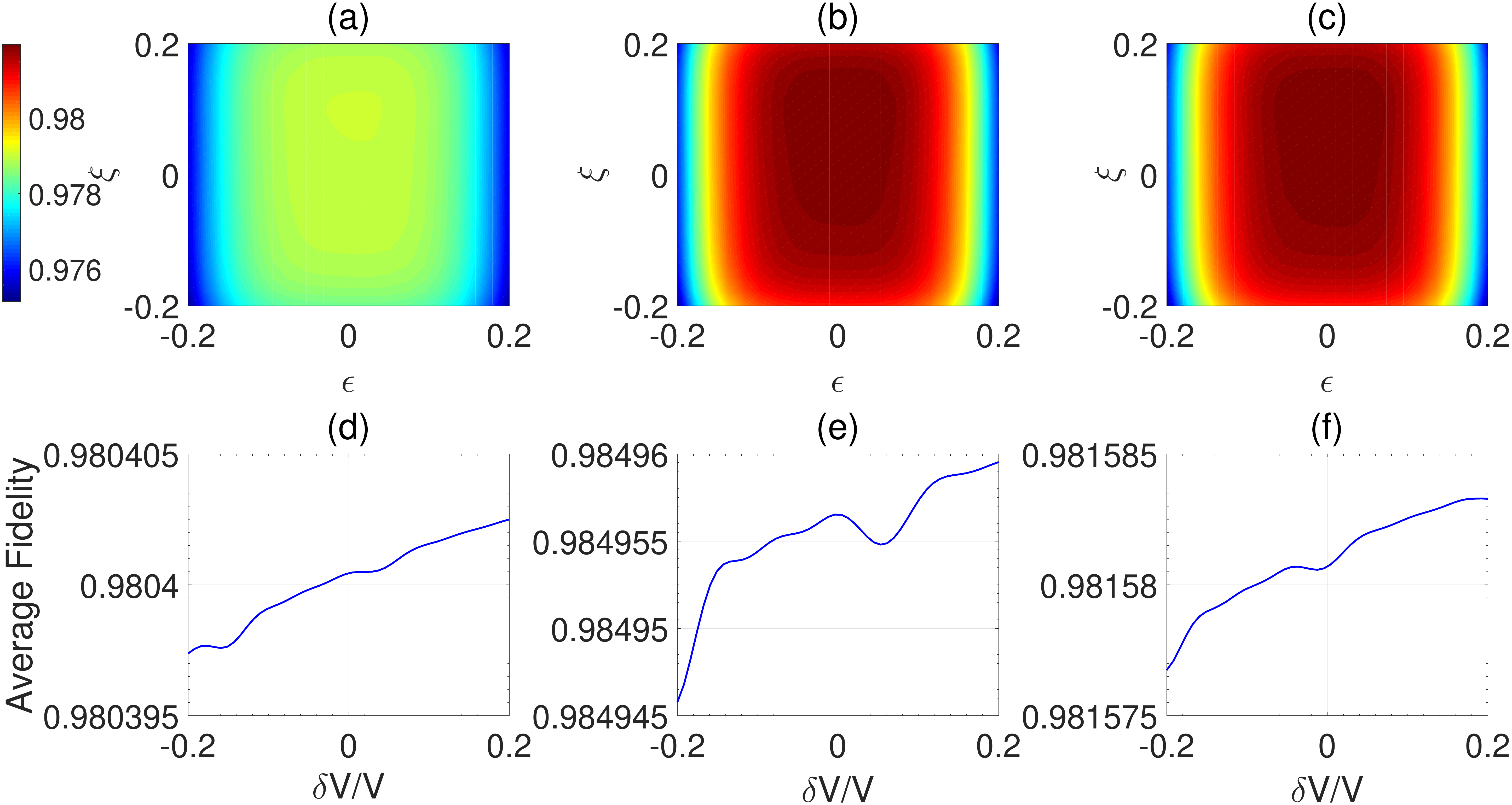}
	\caption{ Average fidelity of CZ (CNOT, and CHadamard) gate versus Rabi error(a)[(b), (c)] and variance of RRI (d)[(e), (f)], respectively. For panels (a), (b) and (c), the variance of RRI is set as 20\%. For panels (d), (e) and (f), Rabi errors for control and target atom are set as 10\%. The maximal Rabi frequency and RRI are chosen based on the experiment~\cite{levine2019parallel}. The other parameters and the pulse shape (here $\tau_1$ and $\tau_2$ should be recalculated through the maximal Rabi frequency) are the same as that of Fig.~\ref{fig2}. }\label{fig13}
\end{figure}

Experimentally, the values of the Rabi frequency as well as the RRI will be lower than the values set and discussed above. We simulate the influence of this case with the consideration of dissipation in Fig.~\ref{fig13} through considering the achieved parameters in experiment~\cite{levine2019parallel} with excitation Rabi frequency from ground to Rydberg state being $2\pi\times3.5$~MHz and RRI being $2\pi\times24$~MHz, respectively. From Fig.~\ref{fig13}(a)[(b), (c)], one can see that, on the premise that the RRI has 20\% fluctuations, the average fidelity of CZ (CNOT and CHadamard) gate are still higher than 0.978(0.98, 0.98), respectively, when the Rabi error of control and target atoms are close to 20\%. Meanwhile, From Fig.~\ref{fig13}(d)[(e), (f)], one can see that, on the premise that the Rabi error for control and target atoms are 10\%, the average fidelity of CZ (CNOT and CHadamard) gate are still higher than 0.98 (0.985, 0.981), respectively, when the RRI has 20\% fluctuations.

\section{Conclusion}\label{sec5}

In conclusion, we have proposed to construct two-bit quantum logic gates with Rydberg atoms based on geometric phase and dark-state dynamics. The results show that, on one hand, the scheme is feasible even when the RRI strength is comparable to the Rabi frequency which may induce strong blockade errors in the conventional blockade scheme. On the other hand, the scheme does not reduce the average fidelity significantly when the control error reaches 10\%. Although the consideration of the motion-induced dephasing with random velocities for individual atoms would decrease the average fidelity, this does not affect the application of our scheme on the Rydberg experimental platform with large Rabi and blockade errors.

\section*{Acknowledgement}

This work was supported by Key Research $\&$ Development Project of Guangdong Province (Grant Nos. 2020B0303300001, 2018B030326001), the National Natural Science Foundation of China (Grant Nos. 12274376, 11835011, 11734018, 11875160), the Natural Science Foundation of Guangdong Province (Grant No. 2017B030308003), the Science, Technology and Innovation Commission of Shenzhen Municipality (Grant Nos. JCYJ20170412152620376, JCYJ20170817105046702, KYTDPT20181011104202253), the Economy, Trade and Information Commission of Shenzhen Municipality (Grant No. 201901161512), and Guangdong Provincial Key Laboratory (Grant No. 2019B121203002). W. L. acknowledges support from the Royal Society through the International Exchanges Cost Share award No. IEC$\backslash$NSFC$\backslash$181078. L.-N. would like to thank J.-Z. Xu for discussions.

\appendix

\section{Derivation of Eq.~(\ref{e10})}\label{appendixA}
As the time-dependent states $|\phi_{k}(t)\rangle$ follows the Schr\"{o}dinger equation, we can write the time evolution operator as,
\begin{equation}\label{s1app}
U(t, 0)=\mathbf{T} e^{-i \int_{0}^{t} H_{c}\left(t^{\prime}\right) d t^{\prime}}=\sum_{k=1,2}\left|\phi_{k}(t)\right\rangle\left\langle\phi_{k}(0)\right| \ ,
\end{equation}
where $\mathbf{T}$ is time ordering operator. On the other hand, the equation between Hamiltonian $H_{c}(t)$ and evolution operator $U(t, 0)$ is given by,
\begin{equation}\label{s2app}
H_{c}(t)=i \dot{U}(t, 0) U^{\dagger}(t, 0) \ .
\end{equation}
Combining Eq.(\ref{s1app}), Eq.(\ref{s2app}) and Eq. (\ref{eq5}), the Hamiltonian $H_{c}(t)$ can be written as
\begin{equation}\label{s3app}
\begin{array}{rlr}
H_{c}(t) & =i \sum_{k=1,2}|\dot{\phi}_{k}(t)\rangle\left\langle\phi_{k}(t)\right| \\
& =\left(\begin{array}{cc}
-\dot{\eta} \sin ^{2} \frac{\theta}{2}-\dot{\gamma} \cos \theta & \frac{1}{2}e^{i \eta}(\dot{i\theta}+M) \\
\frac{1}{2}e^{-i \eta}(-i \dot{\theta}+M) & \dot{\eta} \sin ^{2} \frac{\theta}{2}+\dot{\gamma} \cos \theta
\end{array}\right)
\end{array} \ ,
\end{equation}
with $M=2 \dot{\gamma} \sin \theta-\dot{\eta} \sin \theta$. Note that the Eq.(\ref{s3app}) should be equal to the Eq.(\ref{eq11}), and thus it is not difficult to obtain Eq.~(\ref{e10}).

\begin{widetext}
\section{Demonstration of the parameters in Tabel.~\ref{t1} satisfy the condition in Eq.~(\ref{e9})}\label{appendixB}
\begin{eqnarray}
&&\int_0^{\tau_1}\frac{\dot{\theta}}{2}\exp\left(-i\int_0^{t}\frac{\dot\eta}{\cos\theta}dt' \right)dt\cr\cr
&=&   \int_{0}^{\tau_{1-}/3}\frac{\dot{\theta}}{2}\exp\left(-i\int_0^{t}\frac{\dot\eta}{\cos\theta}dt' \right)dt
+ \int_{\tau_{1-}/3}^{\tau_{1+}/3}\frac{\dot{\theta}}{2}\exp\left(-i\int_0^{t}\frac{\dot\eta}{\cos\theta}dt' \right)dt
+ \int_{\tau_{1+}/3}^{2\tau_{1-}/3}\frac{\dot{\theta}}{2}\exp\left(-i\int_0^{t}\frac{\dot\eta}{\cos\theta}dt' \right)dt\cr\cr
&+& \int_{2\tau_{1-}/3}^{2\tau_{1+}/3}\frac{\dot{\theta}}{2}\exp\left(-i\int_0^{t}\frac{\dot\eta}{\cos\theta}dt' \right)dt
+ \int_{2\tau_{1+}/3}^{\tau_1}\frac{\dot{\theta}}{2}\exp\left(-i\int_0^{t}\frac{\dot\eta}{\cos\theta}dt' \right)dt\cr\cr
&=& \int_{0}^{\tau_{1-}/3}\frac{\dot{\theta}}{2}\exp\left(-i\int_0^{\tau_{1-}/3}\frac{\dot\eta}{\cos\theta}dt' \right)dt + \int_{\tau_{1-}/3}^{\tau_{1+}/3}\frac{\dot{\theta}}{2}\exp\left(-i\int_0^{\tau_{1-}/3}\frac{\dot\eta}{\cos\theta}dt' -i\int_{\tau_{1-}/3}^{\tau_{1+}/3}\frac{\dot\eta}{\cos\theta}dt' \right)dt\cr\cr
&+& \int_{\tau_{1+}/3}^{2\tau_{1-}/3}\frac{\dot{\theta}}{2}\exp\left(-i\int_0^{\tau_{1-}/3}\frac{\dot\eta}{\cos\theta}dt' -i\int_{\tau_{1-}/3}^{\tau_{1+}/3}\frac{\dot\eta}{\cos\theta}dt'-i\int_{\tau_{1+}/3}^{2\tau_{1-}/3}\frac{\dot\eta}{\cos\theta}dt'  \right)dt\cr\cr
&+& \int_{2\tau_{1-}/3}^{2\tau_{1+}/3}\frac{\dot{\theta}}{2}\exp\left(-i\int_0^{\tau_{1-}/3}\frac{\dot\eta}{\cos\theta}dt' -i\int_{\tau_{1-}/3}^{\tau_{1+}/3}\frac{\dot\eta}{\cos\theta}dt'-i\int_{\tau_{1+}/3}^{2\tau_{1-}/3}\frac{\dot\eta}{\cos\theta}dt' -i\int_{2\tau_{1-}/3}^{2\tau_{1+}/3}\frac{\dot\eta}{\cos\theta}dt' \right)dt\cr\cr
&+& \int_{2\tau_{1+}/3}^{\tau_{1}}\frac{\dot{\theta}}{2}\exp\left(-i\int_0^{\tau_{1-}/3}\frac{\dot\eta}{\cos\theta}dt' -i\int_{\tau_{1-}/3}^{\tau_{1+}/3}\frac{\dot\eta}{\cos\theta}dt'-i\int_{\tau_{1+}/3}^{2\tau_{1-}/3}\frac{\dot\eta}{\cos\theta}dt' -i\int_{2\tau_{1-}/3}^{2\tau_{1+}/3}\frac{\dot\eta}{\cos\theta}dt' -i\int_{2\tau_{1+}/3}^{\tau_{1}}\frac{\dot\eta}{\cos\theta}dt' \right)dt\cr\cr
&=& \frac{\pi}{2} + 0 + \frac{\pi}{2}e^{-i\frac{2\pi}{3}} + 0 + \frac{\pi}{2}e^{-i\frac{4\pi}{3}}=0.
\end{eqnarray}
\end{widetext}
And the parameters in Table~\ref{t2} can be demonstrated to satisfy the condition in a similar way.

\section{Adiabatic condition of step (ii)}\label{appendixC}

The adiabatic condition is 
\begin{equation}\label{B1}
	|\langle b_{\pm}|\frac{\partial}{\partial t}|d\rangle_2|\ll|\pm\mathcal{N}-0|
\end{equation}
The left hand of Eq.~(\ref{B1}) can be calculated as 
\begin{equation}\label{B2}
|\langle b_{\pm}|\frac{\partial}{\partial t}|d\rangle_2|=\frac{|\sqrt{2}V\frac{\partial\Omega_{t}^*}{\partial t}|}{4\mathcal{N}^2}
\end{equation}
Thus, Eq.~(\ref{B1}) can be simplified as 
\begin{equation}
\left|\frac{\partial\Omega_{t}^*}{\partial t}\right|\ll\frac{2\sqrt{2}\mathcal{N}^3}{V}
\end{equation}
We can choose parameters to make $|\Omega_t|\dot{\varphi}_2\equiv0$~(In fact, the parameters in Table.~\ref{t2} satisfy this condition).
That is the variation rate of the absolute value of Rabi frequency should satisfy 
\begin{equation}
\frac{\partial\left|\Omega_{t}\right|}{\partial t}\ll\frac{2\sqrt{2}\mathcal{N}^3}{V}.
\end{equation}

\section{Derivation of Eqs.~(\ref{eq13}) and (\ref{eq14})}\label{appendixD}
For Eq.~(\ref{eq13}), since $|d\rangle_2$ is the dark state of $H_{ii}$, one can get $H_{ii}|d\rangle_2 = 0$ and thus $\varphi_{dy}=0$.
For Eq.~(\ref{eq14}),
\begin{widetext}
\begin{eqnarray}\label{a1}
\int_{\tau_1}^{\tau_1+\tau_2} \frac{\Omega_{t}^2 \dot{\varphi}_{2}}{\Omega_{t}^2+4 V^2}dt =&& \int_{\tau_1}^{\tau_1+\frac{\tau_{2-}}{4}} \frac{\Omega_{t}^2 \dot{\varphi}_{2}}{\Omega_{t}^2+4 V^2}dt
+ \int_{\tau_1+\frac{\tau_{2-}}{4}}^{\tau_1+\frac{\tau_{2+}}{4}} \frac{\Omega_{t}^2 \dot{\varphi}_{2}}{\Omega_{t}^2+4 V^2}dt +  \int_{\tau_1+\frac{\tau_{2+}}{4}}^{\tau_1+\frac{\tau_{2-}}{2}} \frac{\Omega_{t}^2 \dot{\varphi}_{2}}{\Omega_{t}^2+4 V^2}dt +  \int_{\tau_1+\frac{\tau_{2-}}{2}}^{\tau_1+\frac{\tau_{2+}}{2}} \frac{\Omega_{t}^2 \dot{\varphi}_{2}}{\Omega_{t}^2+4 V^2}dt \cr\cr &&
+  \int_{\tau_1+\frac{\tau_{2+}}{2}}^{\tau_1+\frac{3\tau_{2-}}{4}} \frac{\Omega_{t}^2 \dot{\varphi}_{2}}{\Omega_{t}^2+4 V^2}dt +  \int_{\tau_1+\frac{3\tau_{2-}}{4}}^{\tau_1+\frac{3\tau_{2+}}{4}} \frac{\Omega_{t}^2 \dot{\varphi}_{2}}{\Omega_{t}^2+4 V^2}dt +  \int_{\tau_1+\frac{3\tau_{2+}}{4}}^{\tau_1+\tau_{2}} \frac{\Omega_{t}^2 \dot{\varphi}_{2}}{\Omega_{t}^2+4 V^2}dt\cr\cr=&&
\int_{\tau_1}^{\tau_1+\frac{\tau_{2-}}{4}} \frac{\Omega_{t}^2 \times0}{\Omega_{t}^2+4 V^2}dt
+ \int_{\tau_1+\frac{\tau_{2-}}{4}}^{\tau_1+\frac{\tau_{2+}}{4}} \frac{0\times \dot{\varphi}_{2}}{0+4 V^2}dt +  \int_{\tau_1+\frac{\tau_{2+}}{4}}^{\tau_1+\frac{\tau_{2-}}{2}} \frac{\Omega_{t}^2 \times 0}{\Omega_{t}^2+4 V^2}dt +  \int_{\tau_1+\frac{\tau_{2-}}{2}}^{\tau_1+\frac{\tau_{2+}}{2}} \frac{0\times \dot{\varphi}_{2}}{0+4 V^2}dt \cr\cr &&
+  \int_{\tau_1+\frac{\tau_{2+}}{2}}^{\tau_1+\frac{3\tau_{2-}}{4}} \frac{\Omega_{t}^2 \times0}{\Omega_{t}^2+4 V^2}dt +  \int_{\tau_1+\frac{3\tau_{2-}}{4}}^{\tau_1+\frac{3\tau_{2+}}{4}} \frac{0\times \dot{\varphi}_{2}}{0+4 V^2}dt +  \int_{\tau_1+\frac{3\tau_{2+}}{4}}^{\tau_1+\tau_{2}} \frac{\Omega_{t}^2 \times 0}{\Omega_{t}^2+4 V^2}dt\cr\cr=&& 0.
\end{eqnarray}
\end{widetext}

\section{Re-derivation of the geometric phase in Eq.~(\ref{eq14}) when V is time-dependent}\label{appendixE}
In practical case, $V$ is related to interatomic distance $d$ and $d$ is linearly related to time $t$. We set $d = a*t$ where $a$ is a time-independent parameter. Thus we can get $V(t) = C_3/(a^3t^3)$. On that basis, we calculate the geometric phase as 
\begin{equation}\label{a2}
\varphi_{ge}=i\int_{\tau_1}^{\tau_1+\tau_2}~_{2}\langle d|\frac{\partial}{\partial t}|d\rangle_2dt=\int_{\tau_1}^{\tau_1+\tau_2} \frac{a^6t^6\Omega_{t}^2 \dot{\varphi}_{2}}{a^6t^6\Omega_{t}^2+4(C_3)^2}dt.
\end{equation}
After considering the concrete pulses, the geometric phase is calculated as
\begin{widetext}
	\begin{eqnarray}\label{a3}
\varphi_{ge} =&& \int_{\tau_1}^{\tau_1+\frac{\tau_{2-}}{4}} \frac{a^6t^6\Omega_{t}^2 \dot{\varphi}_{2}}{a^6t^6\Omega_{t}^2+4 (C_3)^2}dt
	+ \int_{\tau_1+\frac{\tau_{2-}}{4}}^{\tau_1+\frac{\tau_{2+}}{4}} \frac{a^6t^6\Omega_{t}^2 \dot{\varphi}_{2}}{a^6t^6\Omega_{t}^2+4 (C_3)^2}dt +  \int_{\tau_1+\frac{\tau_{2+}}{4}}^{\tau_1+\frac{\tau_{2-}}{2}} \frac{a^6t^6\Omega_{t}^2 \dot{\varphi}_{2}}{a^6t^6\Omega_{t}^2+4 (C_3)^2}dt +  \int_{\tau_1+\frac{\tau_{2-}}{2}}^{\tau_1+\frac{\tau_{2+}}{2}} \frac{a^6t^6\Omega_{t}^2 \dot{\varphi}_{2}}{a^6t^6\Omega_{t}^2+4 (C_3)^2}dt \cr\cr &&
	+  \int_{\tau_1+\frac{\tau_{2+}}{2}}^{\tau_1+\frac{3\tau_{2-}}{4}} \frac{a^6t^6\Omega_{t}^2 \dot{\varphi}_{2}}{a^6t^6\Omega_{t}^2+4 (C_3)^2}dt +  \int_{\tau_1+\frac{3\tau_{2-}}{4}}^{\tau_1+\frac{3\tau_{2+}}{4}} \frac{a^6t^6\Omega_{t}^2 \dot{\varphi}_{2}}{a^6t^6\Omega_{t}^2+4 (C_3)^2}dt +  \int_{\tau_1+\frac{3\tau_{2+}}{4}}^{\tau_1+\tau_{2}} \frac{a^6t^6\Omega_{t}^2 \dot{\varphi}_{2}}{a^6t^6\Omega_{t}^2+4 (C_3)^2}dt\cr\cr=&&
	\int_{\tau_1}^{\tau_1+\frac{\tau_{2-}}{4}} \frac{a^6t^6\Omega_{t}^2 \times0}{a^6t^6\Omega_{t}^2+4 (C_3)^2}dt
	+ \int_{\tau_1+\frac{\tau_{2-}}{4}}^{\tau_1+\frac{\tau_{2+}}{4}} \frac{0\times \dot{\varphi}_{2}}{0+4 (C_3)^2}dt +  \int_{\tau_1+\frac{\tau_{2+}}{4}}^{\tau_1+\frac{\tau_{2-}}{2}} \frac{a^6t^6\Omega_{t}^2 \times 0}{a^6t^6\Omega_{t}^2+4 (C_3)^2}dt +  \int_{\tau_1+\frac{\tau_{2-}}{2}}^{\tau_1+\frac{\tau_{2+}}{2}} \frac{0\times \dot{\varphi}_{2}}{0+4 (C_3)^2}dt \cr\cr &&
	+  \int_{\tau_1+\frac{\tau_{2+}}{2}}^{\tau_1+\frac{3\tau_{2-}}{4}} \frac{a^6t^6\Omega_{t}^2 \times0}{a^6t^6\Omega_{t}^2+4 (C_3)^2}dt +  \int_{\tau_1+\frac{3\tau_{2-}}{4}}^{\tau_1+\frac{3\tau_{2+}}{4}} \frac{0\times \dot{\varphi}_{2}}{0+4 (C_3)^2}dt +  \int_{\tau_1+\frac{3\tau_{2+}}{4}}^{\tau_1+\tau_{2}} \frac{a^6t^6\Omega_{t}^2 \times 0}{a^6t^6\Omega_{t}^2+4 (C_3)^2}dt\cr\cr=&&~ 0.
	\end{eqnarray}
\end{widetext}

\section{Approximated excitation error of a single channel}\label{af}
For simplicity, we treat all of the possible imperfection excitation process as a series of two-level systems constituted by the any one of leakage states and the ground state. The Hamiltonian can be written as 
\begin{equation}
    H_{\rm leakage} = \delta|R_L\rangle\langle R_L| + \Omega/2(|g\rangle\langle R_L| + {\rm H.c.}),
\end{equation}
where $\delta$ denotes the detuning and $\Omega$ the Rabi frequency, $|g\rangle$ is the ground state ($|0\rangle$ or $|1\rangle$) and $|R_L\rangle$ is the leakage state. 
One can get the population of the state $|R_L\rangle$ 
\begin{equation}\label{f2}
P_{R_L}= |\frac{i\Omega e^{-\frac{1}{2} i\delta  t} \sin \left(\frac{1}{2} t \sqrt{\delta ^2+\Omega ^2}\right)}{\sqrt{\delta ^2+\Omega ^2}}|^2.  
\end{equation}
For simplicity, we can get the series expansion by considering the condition $\Omega/\delta\ll1$ and one $\pi$ pulse time for the resonant case $t = \pi/\Omega$. Then, Eq.~(\ref{f2}) can be simplified as (we set $m=\Omega/\delta$ for simplicity)
\begin{widetext}
\begin{equation}\label{f3}
P_{R_L}\approx\left(m^2-m^4+m^6+O\left(m^7\right)\right) \sin ^2\left(\frac{1}{2} \sqrt{\frac{1}{m^2}} \pi +\frac{1}{4} \sqrt{\frac{1}{m^2}} \pi  m^2-\frac{1}{16} \left(\sqrt{\frac{1}{m^2}} \pi \right) m^4+\frac{1}{32} \sqrt{\frac{1}{m^2}} \pi  m^6+O\left(m^8\right)\right).
\end{equation}
\end{widetext}
Since the maximal value of $(\sin)^2$ is 1 and $m^2\ll1$, the following relationship can be derived from Eq.~(\ref{f3}) as
\begin{equation}\label{f4}
P_{R_L} \leq m^2 = (\Omega/\delta)^2.  
\end{equation}

\bibliography{REV}

\end{document}